\documentclass[twocolumn,pra]{revtex4}
\usepackage[utf8]{inputenc}
\usepackage{amsmath}
\usepackage{amssymb}
\usepackage{bm}
\usepackage{graphicx}
\usepackage{color}
\DeclareMathOperator\arccosh{arccosh}

\newcommand{\cL}{\mathcal{L}}
\newcommand{\sign}{\mathop{\mathrm{sign}}\nolimits}
\renewcommand{\Im}{\mathop{\mathrm{Im}}\nolimits}
\renewcommand{\Re}{\mathop{\mathrm{Re}}\nolimits}
\newcommand\blfootnote[1]{%
  \begingroup
  \renewcommand\thefootnote{}\footnote{#1}%
  \addtocounter{footnote}{-1}%
  \endgroup
}

\begin{document}

\title{Theory of coherent quantum phase-slips in Josephson junction chains with periodic spatial modulations}

\author{Aleksandr E. Svetogorov,$^{1,2,3}$ Masahiko Taguchi,$^{4}$ Yasuhiro Tokura,$^{5}$ Denis M. Basko,$^{1}$ Frank W. J. Hekking$^{1}\dagger$\blfootnote{$\dagger$ Deceased 15 May 2017.}}
\affiliation{
$^{1}$Laboratoire de Physique et Mod\'elisation des Milieux Condens\'es,
Universit\'e Grenoble Alpes and CNRS,
25 rue des Martyrs, 38042 Grenoble, France}
\affiliation{$^{2}$Landau Institute for Theoretical Physics, Russian Academy of Sciences, 142432 Chernogolovka, Russia}
\affiliation{$^{3}$Moscow Institute of Physics and Technology, 141700, Dolgoprudny, Russia}
\affiliation{$^{4}$Graduate School of Science, Kyoto University, Kyoto 606-8502, Japan}
\affiliation{$^{5}$Graduate School of Pure and Applied Sciences, University of Tsukuba,
1-1-1 Tennodai, Tsukuba, Ibaraki 305-8571, Japan}

\footnotetext{Deceased 15 May 2017}

\date{\today}
\begin{abstract}
We study coherent quantum phase-slips which lift the ground state degeneracy in a Josephson junction ring, pierced by a magnetic flux of the magnitude equal to half of a flux quantum. The quantum phase-slip amplitude is sensitive to the normal mode structure of superconducting phase oscillations in the ring (Mooij-Sch\"on modes). These, in turn, are affected by spatial inhomogeneities in the ring. We analyze the case of weak periodic modulations of the system parameters and calculate the corresponding modification of the quantum phase-slip amplitude.
\end{abstract}

\maketitle

\section{Introduction}

Peculiarities of superconductivity in one-dimensional systems attracted interest of the scientific community long ago~\cite{Frohlich1954, Ferrell1964, Little1964, Rice1965, Bychkov1966}. Since then, an important role played by phase-slips has been realized~\cite{Little1967}. A quantum phase-slip (QPS) is a sudden change of the superconducting phase difference along a one-dimensional superconductor by $2\pi$ via quantum-mechanical tunneling. Presently, one-dimensional superconductivity can be realized in Josephson junction (JJ) chains or thin metallic wires (see Refs.~\cite{Fazio2001} and~\cite{Arutyunov2008} for respective reviews). In a good superconductor the QPSs have low probability, but they can give rise to qualitatively new effects, such as small but finite dc resistance of the superconductor at low temperatures~\cite{Giordano1988, Bezryadin2000, Lau2001, Altomare2006, Zgirski2008}, or system coupling to external charges~\cite{Ivanov2001, Matveev2002, Hriscu2011PRB, Manucharyan2012, Pop2012}. If the QPSs become frequent enough, they can turn the system into an insulator~\cite{Bradley1984, Korshunov1986, Korshunov1989, Meidan2007}. One distinguishes between incoherent QPSs, which are accompanied by energy dissipation, and coherent QPSs, which only shift the system energy levels. Both incoherent \cite{Giordano1988, Lau2001, Altomare2006, Zgirski2008, Sahu2009, Pop2010, Hongisto2012, Arutyunov2012} and coherent~\cite{Manucharyan2009, Manucharyan2012, Astafiev2012, Peltonen2013} QPSs have been observed experimentally. On the one hand, QPSs are of fundamental interest as they manifest the quantum behavior of a macroscopic collective degree of freedom (the superconducting phase). On the other, new devices based on coherent QPSs have been proposed~\cite{Mooij2005, Mooij2006, Guichard2010, Hriscu2011PRL}.

When phase tunneling in a JJ chain is described quasiclassically~\cite{Matveev2002, Rastelli2013}, the amplitude of a single QPS is proportional to $e^{-S_\mathrm{QPS}}$, where $S_\mathrm{QPS}\gg{1}$ is the action on the classical imaginary-time trajectory corresponding to the QPS (we set $\hbar=1$ throughout the paper). This classical trajectory involves phase winding by $2\pi$ on one of the junctions, accompanied by phase readjustment to the slipped configuration in the rest of the chain. This readjustment is governed by the gapless Mooij-Sch\"on modes~\cite{Dahm1968, Kulik1974, Mooij1985, Camarota2001, Pop2011, Masluk2012}, which play the role of the environment for the QPS. This environment contribution diverges logarithmically with the chain length $L$ and gives rise to the logarithmic interaction between phase-slips in multi-QPS configurations~\cite{Bradley1984, Korshunov1986, Korshunov1989}. 

Here we study spatially inhomogeneous JJ chains. One can distinguish two types of inhomogeneities. One type correponds to inhomogeneous charge distribution along the system (which may be due to extrinsic static charges, inhomogeneous external potentials, electron density modulations, etc.); this results in different phases of the QPS amplitudes at different junctions due to Aharonov-Casher effect~\cite{Ivanov2001, Matveev2002}, but does not change~$S_\mathrm{QPS}$. Random inhomogeneities of this kind were the primary interest in Ref.~\cite{Bard2017}. The second type is an inhomogeneity of the island or junction sizes along the chain, which is our main interest in this paper. An extreme case is when one junction is much smaller than all the rest, so the QPS is pinned to this junction; this situation was analyzed in Refs.~\cite{Hekking1997, Vanevic2012}. However, the environment part of the action is not affected in this situation, and we are not aware of any calculation of $S_\mathrm{QPS}$ in a spatially inhomogeneous system.

Obviously, the Mooij-Sch\"on modes are affected by spatial modulations. In the disordered case, they are all localized~\cite{Basko2013}. The effect of periodic spatial modulations on the Debye-Waller factor of Mooij-Sch\"on modes has been studied in Ref.~\cite{Taguchi2015}. Here we analyze the QPS action in a JJ chain whose parameters have a weak periodic spatial modulation. We calculate the correction to $S_\mathrm{QPS}$ due to the modulation, and find that it has the same form as the main term, but the logarithmic divergence is cut off at the modulation period rather than the system length. The case of random modulations will be the subject of a future study. As a by-product of our calculation, we obtain a more precise expression for $S_\mathrm{QPS}$ in a homogeneous JJ chain than the one given in Ref.~\cite{Rastelli2013}.

Although here we focus on JJ chains, their low-frequency properties are quite similar to those of thin superconducting wires (in fact, it was a weak link in a superconducting wire that was studied in Refs.~\cite{Hekking1997, Vanevic2012, Taguchi2015}). The logarithmic part of the QPS action is determined by the Mooij-Sch\"on modes with low frequencies and is the same for wires and JJ chains. Our results are applicable to spatially modulated wires as well.

This paper is organized as follows. In Sec.~\ref{sec:model} we specify the model, briefly review the previously known facts about spatially homogeneous JJ chains, and summarize our results for the periodically modulated system. In Sec.~\ref{sec:general} we derive the general expressions for the QPS action in a spatially inhomogeneous system. The specific case of the periodic modulation is addressed in Sec.~\ref{sec:periodic}. In Sec.~\ref{sec:wires} we discuss applicability of our results to superconducting wires. In Sec.~\ref{sec:conclusions} we give our conclusions. In Appendix~\ref{app:homogeneous} we give a calculation of the QPS amplitude in homogeneous JJ chains.

\section{Model, qualitative discussion, and summary of results}
\label{sec:model}

\subsection{Phase action for Josephson junction chains}
\label{ssec:wires-junctions}

\begin{figure}
\includegraphics[width=7cm]{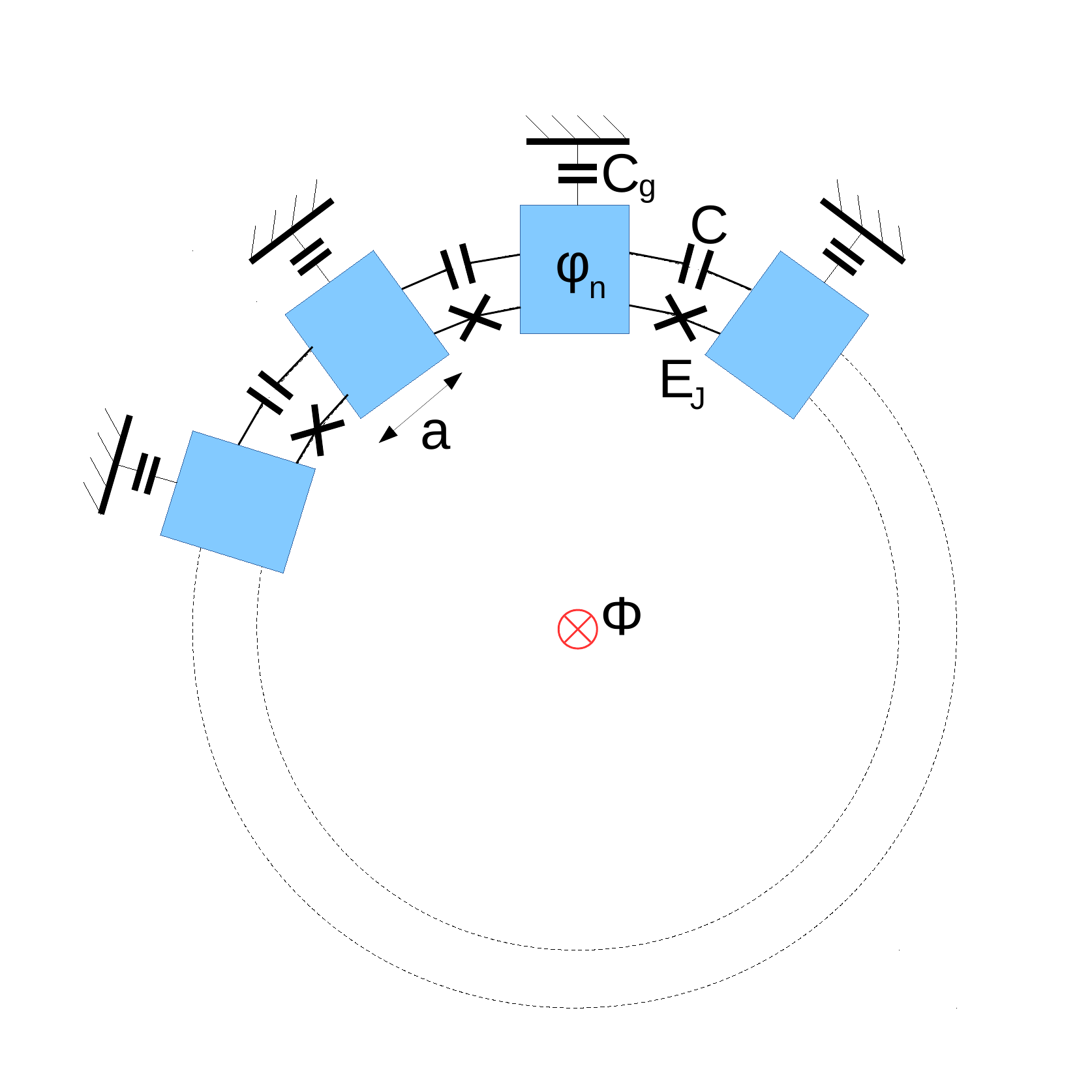}
\caption{\label{fig:JJchain}
A schematic representation of a superconducting ring
threaded by a magnetic flux $\Phi$ and containing $N$ Josephson
junctions with a capacitance $C$ between the neighbouring islands and a capacitance $C_g$ to the ground. $E_J$ is the Josephson energy, $a$ is the size of each superconducting island. $\phi_n$ is the condensate phase of the $n$th superconducting island.}
\end{figure}

The model system is implemented as a chain of Josephson junctions between neighboring superconducting islands. Then, the dynamical variables are the phases $\phi_n$, where the integer $n$ labels the islands. We assume that the chain of $N+1$ junctions is closed in a ring, pierced by a magnetic flux (Fig. \ref{fig:JJchain}). Then, the island $n=0$ is identified with the island $n=N+1$, so that $\phi_0=\phi_{N+1}$, which corresponds to periodic boundary conditions. 

Since we are going to study quantum tunnelling of the superconducting phase $\phi$ in the quasiclassical limit, it is natural to pass to imaginary time $\tau$ and describe the system by its zero-temperature Euclidean action,
which can be written as~\cite{Fazio2001}
\begin{multline}\label{eq:SJJ}
S=\int{d}\tau\sum_{n=0}^{N}\Big[ \frac{C_\mathrm{g}}{8e^{2}}\,\dot{\phi}_n^2+\frac{C}{8e^{2}}
\left(\dot{\phi}_{n+1}-\dot{\phi}_{n}\right)^2\\
{}-E_{J}\cos\left(\phi_{n+1}-\phi_{n}+\Phi/(N+1)\right)\Big],
\end{multline}
where $\dot\phi_n\equiv\partial\phi_n/\partial\tau$. Here $E_J$ and $C$ are the Josephson energy and the capacitance of the junction between two neighbouring islands, while $C_\mathrm{g}$ is the capacitance between an island and a nearby ground plane; $\Phi$ is the magnetic flux in units of the superconducting flux quantum divided by $2\pi$ (one flux quantum piercing the ring corresponds to $\Phi=2\pi$).
Typically, $C\gg{C}_\mathrm{g}$~\cite{Pop2011, Masluk2012}. We also assume $E_J\gg{e}^2/C$.

In principle, the phase can slip on any of the $N$ junctions; QPSs at different junctions contribute to the same quantum transition (i.~e., with the same initial and final states), so their amplitudes should be added up coherently~\cite{Ivanov2001, Matveev2002}. Let us choose one of the junctions and study the corresponding amplitude. It is also convenient to number the junctions so that the slipping junction is the one between the islands $n=N$ and $n=0$, which we will call ``boundary''. Then, it is convenient to perform a gauge transformation,
\begin{equation}
\phi_n\to\phi_n-\frac{n}{N+1}\,\Phi,
\end{equation}
which corresponds to twisted boundary conditions, $\phi_{N+1}=\phi_0+\Phi$. Then the flux disappears from all cosine terms in Eq.~(\ref{eq:SJJ}), except the last one, which becomes $-E_J\cos(\phi_N-\phi_0-\Phi)$.

The coefficients $C_\mathrm{g},C,E_J$ in Eq.~(\ref{eq:SJJ}) need not be the same for all junctions and islands. It is possible to fabricate Josephson junction chains with different parameters for different junctions; in principle, an arbitrary spatial pattern can be produced.
One example is when one of the junctions is much smaller than the rest, then the QPS amplitude on this junction dominates over the rest~\cite{Manucharyan2009, Ergul2013, Ergul2017}.
To describe this situation, we introduce the explicit notations $\tilde{C}$ and $\tilde{E}_J$ for the capacitance and the Josephson energy of the boundary junction between $n=N$ and $n=0$. We will analyze the general situation when the relation between $\tilde{C},\tilde{E}_J$ and $C,E_J$ can be arbitrary, including also the special case $\tilde{C}\ll{C}$, $\tilde{E}_J\ll{E}_J$.

\begin{figure}
\includegraphics[width=8cm]{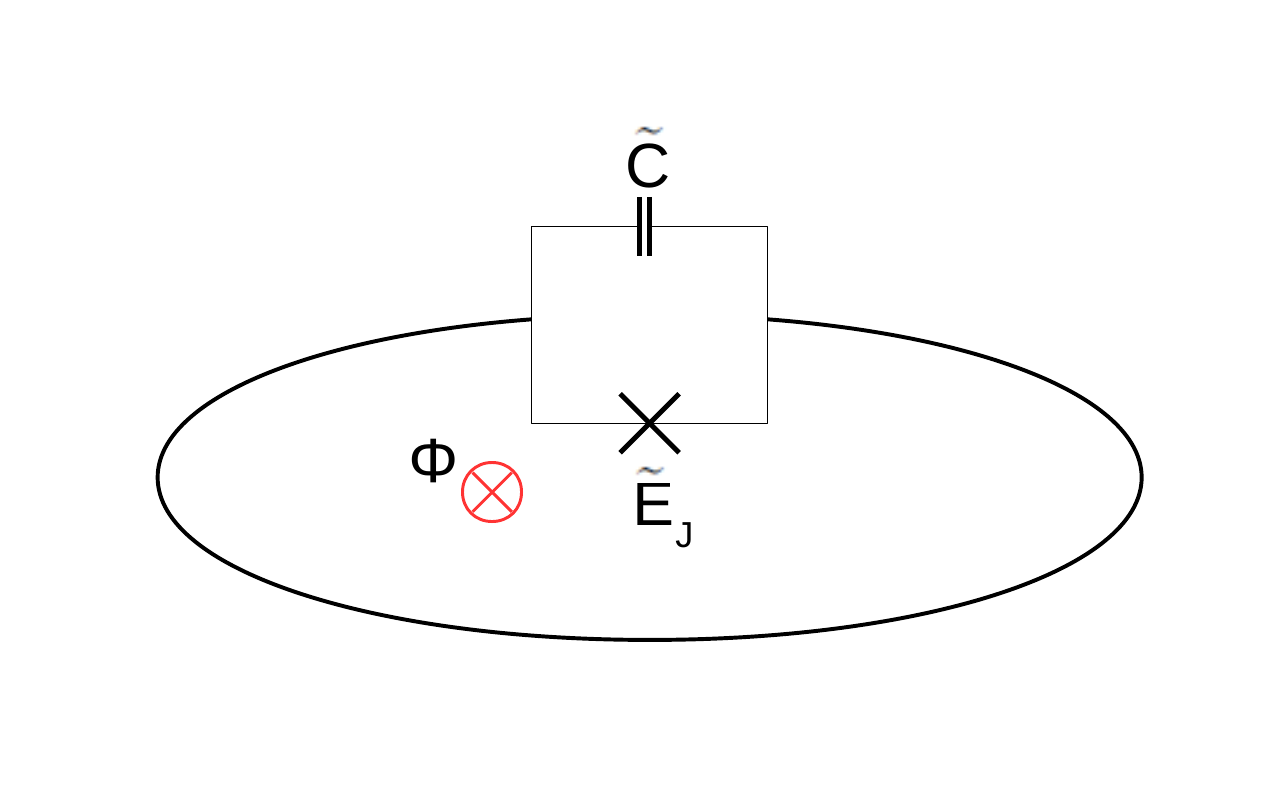}
\caption{\label{fig:Continuum}
A schematic representation of the Josephson junction chain in the continuum limit with the boundary junction shown explicitly.
}
\end{figure}

While allowing one junction in the chain to be strongly different from others, for the rest of the junctions we assume spatial modulations of the parameters to be smooth on the length scale of the island size~$a$, which plays the role of the lattice constant. Then it is convenient to 
pass to the continuum limit for the rest of the junctions, treating the bulk of the ring and the boundary junction separately (Fig. \ref{fig:Continuum}). Namely, we take the limit $na\to{x}$, $\phi_n\to\phi(x)$, $\phi_{n+1}-\phi_n\to a(\partial\phi/\partial{x})$, $\sum_n\to\int{d}x/a$, $Na=L$, and the action can be written as
\begin{align}\label{eq:S_wire+junction}
&S=\int{d}\tau\int_{0}^{L}dx\,\cL(\phi,\partial_\tau\phi)\nonumber\\
&\qquad{}+\int\frac{d\tau}{2\tilde{E}_C}\left[
\frac{\partial\phi(L,\tau)}{\partial\tau}
-\frac{\partial\phi(0,\tau)}{\partial\tau}\right]^2\nonumber\\
&\qquad{}-\int{d}\tau\,\tilde{E}_{J}\cos[\phi(L,\tau)-\phi(0,\tau)-\Phi],\\
&\cL=\frac{1}{2e_c}
\left(\frac{\partial\phi}{\partial\tau}\right)^{2}
+\frac{\ell_s^2}{2e_c}
\left(\frac{\partial^2\phi}{\partial{x}\,\partial\tau}\right)^{2}
+\frac{e_l}{2}
\left(\frac{\partial\phi}{\partial x}\right)^{2}.
\label{eq:loopL}
\end{align}
Here we denoted $e_c=a(2e)^2/C_\mathrm{g}$, $\ell_s^2=a^2C/C_\mathrm{g}$, $e_l=aE_J$, $\tilde{E}_C=(2e)^2/\tilde{C}$. Here $\ell_s\gtrsim{a}$ represents the screening length for the electrostatic Coulomb interaction between charges on different islands. While the lattice action~(\ref{eq:SJJ}) describes the Josephson junction chain down to the shortest length scale, the island size~$a$, action (\ref{eq:S_wire+junction}) is coarse-grained. Still, it turns out to be well-behaved at short distances due to the second term in $\cL$ with the mixed derivatives; no extrinsic ultraviolet cutoff will be needed in the theory. We only assume that the coefficients $1/e_c,\ell_s^2/e_c,e_l$ smoothly depend on~$x$.

It is convenient to characterize the chain by its low-frequency impedance in the units of superconducting conductance quantum $(2e)^2/(\pi\hbar)$ (we momentarily restore~$\hbar$), or its inverse, the dimensionless admittance:
\begin{equation}
g\equiv\pi\sqrt{e_l/e_c}.
\end{equation}
In the following we assume $g\gtrsim{1}$, otherwise the chain would be in the insulating rather then the superconducting state~\cite{Bradley1984, Korshunov1986, Korshunov1989}.

\subsection{Normal modes}

\begin{figure}
\includegraphics[width=8cm]{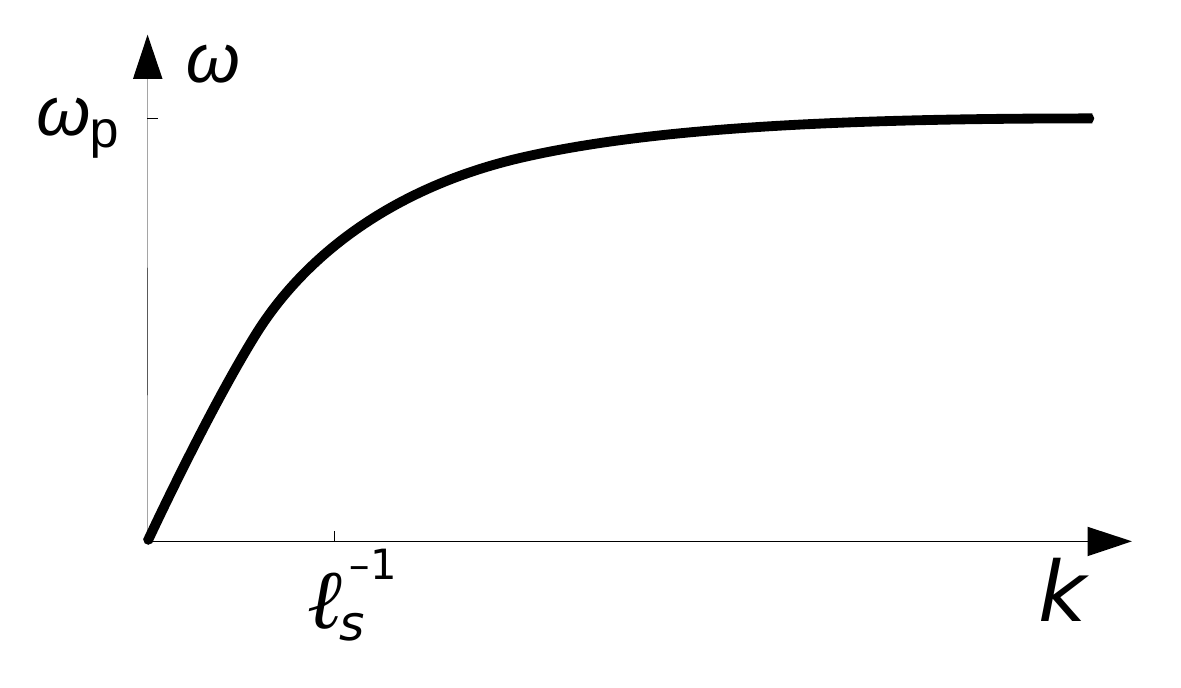}
\caption{\label{fig:dispersion}
The dispersion curve of the phase oscillations (plasmons, Mooij-Sch\"on modes), determined by Eq.~(\ref{eq:dispersion}).
}
\end{figure}

To deal with the slow phase dynamics described by the quadratic Lagrangian density~(\ref{eq:loopL}), it is convenient to decompose the phase field $\phi(x)$ into the normal modes. Namely, we write down the Euler-Lagrange equations of motion and look for the solutions in the form $\phi(x,\tau)=\Psi(x)\,e^{\pm\omega\tau}$ (since $\tau$ is the imaginary time). This gives the following equation for the normal mode wave functions:
\begin{equation}\label{eq:Helmholtz}
\frac{\omega^2}{e_c}\,\Psi
-\omega^2\,\frac\partial{\partial{x}}\frac{\ell_s^2}{e_c}\frac{\partial\Psi}{\partial{x}}
+\frac\partial{\partial{x}}\,e_l\,\frac{\partial\Psi}{\partial{x}}=0,
\end{equation}
with the Dirichlet boundary conditions, $\Psi(0)=\Psi(L)=0$ at $x=0,L$, as will be discussed in detail in Sec.~\ref{ssec:variables}.

For a spatially homogeneous chain, the solutions are plane waves, $\Psi(x)\propto\sin{kx}$, for which Eq.~(\ref{eq:Helmholtz}) gives the dispersion relation~\cite{Pop2011, Masluk2012}:
\begin{equation}\label{eq:dispersion}
\omega(k)=\frac{\omega_\mathrm{p}|k|\ell_s}{\sqrt{1+k^2\ell_s^2}},
\end{equation}
where $\omega_\mathrm{p}=\sqrt{e_ce_l}/\ell_s$ is the plasma frequency of a single junction in the chain. At small $|k|\ll{1}/\ell_s$ the dispersion is linear, $\omega\approx{v}_\mathrm{p}|k|$, characterized by the plasma velocity $v_\mathrm{p}=\sqrt{e_ce_l}$~\cite{Mooij1985}.
In the limit $L\to\infty$, modes with frequencies $\omega\ll\omega_\mathrm{p}$ effectively form an Ohmic bath.
In the following, much of the effort in calculating the QPS amplitude for a modulated JJ chain will be dedicated to solving Eq.~(\ref{eq:Helmholtz}) with space-dependent coefficients which describes the modification of the bath properties by the spatial modulation.

\subsection{Classical phase configurations}
\label{ssec:classicalconf}

For each value of $\Phi$, there is a single static classical phase configuration, minimizing the potential energy. The exception is for $\Phi$ being an odd multiple of~$\pi$, when there are two configurations with equal potential energies. Quantum tunnelling between these degenerate configurations is the main subject of our study. As the dependence of action~(\ref{eq:S_wire+junction}) on $\Phi$ is periodic, we can focus on $\Phi=\pi$ without loss of generality.

Let us find the classical phase configurations taking into account the spatial dependence $e_l(x)$. Minimization of the bulk action leads to the equation
\begin{equation}
\frac\partial{\partial{x}}\,\frac{\partial\cL}{\partial\phi}=
\frac\partial{\partial{x}}\,e_l\,\frac{\partial\phi}{\partial{x}}=0,
\end{equation}
which is nothing but the current conservation. Its solution contains two integration constants, $\phi_0$ and $\vartheta$:
\begin{equation}
\phi(x)=\phi_0 +
\vartheta\,\frac{\int_0^xe_l^{-1}(x')\,dx'}{\int_0^Le_l^{-1}(x')\,dx'}.
\end{equation}
The constant $\vartheta$ should be found by minimizing the total potential energy including the boundary term~\cite{Hekking1997}:
\begin{equation}\label{eq:minphistar}
\frac\partial{\partial\vartheta}\left[
\frac{\vartheta^2/2}{\int_0^Le_l^{-1}(x)\,dx}
-\tilde{E}_J\cos(\vartheta-\Phi)\right]=0.
\end{equation}
At this point it is convenient to introduce the length scale
\begin{equation}
\ell_J\equiv\frac{L}{\tilde{E}_J\int_0^Le_l^{-1}(x)\,dx}.
\end{equation}
As the integral in the denominator is proportional to $L$, the scale $\ell_J$ does not depend on the chain length. 
For a spatially homogeneous chain, $\ell_J/a$ is the number of chain junctions which has the same Josephson inductance as that corresponding to $\tilde{E}_J$. If all $E_J=\tilde{E}_J$, then $\ell_J=a$; for $E_J\gg\tilde{E}_J$, one has $\ell_J\gg{a}$. For a spatially inhomogeneous chain, it is the spatial average of the inductance that enters. Then, Eq.~(\ref{eq:minphistar}) can be written as~\cite{Hekking1997}
\begin{equation}
\frac{\ell_J}L\,\vartheta+\sin(\vartheta-\Phi)=0.
\end{equation}
We assume $L\gg\ell_J$, then $\vartheta\approx\Phi+2\pi{m}$ with any integer $m$ gives a local minimum (half-integer values of $m$ give local maxima) with the potential energy $(\Phi+2\pi{m})^2\tilde{E}_J\ell_J/(2L)$. If $\Phi=\pi$, then the two configurations with $\vartheta=\pi$ and $\vartheta=-\pi$ have the same energies. 

\begin{figure}[h]
\includegraphics[width=8cm]{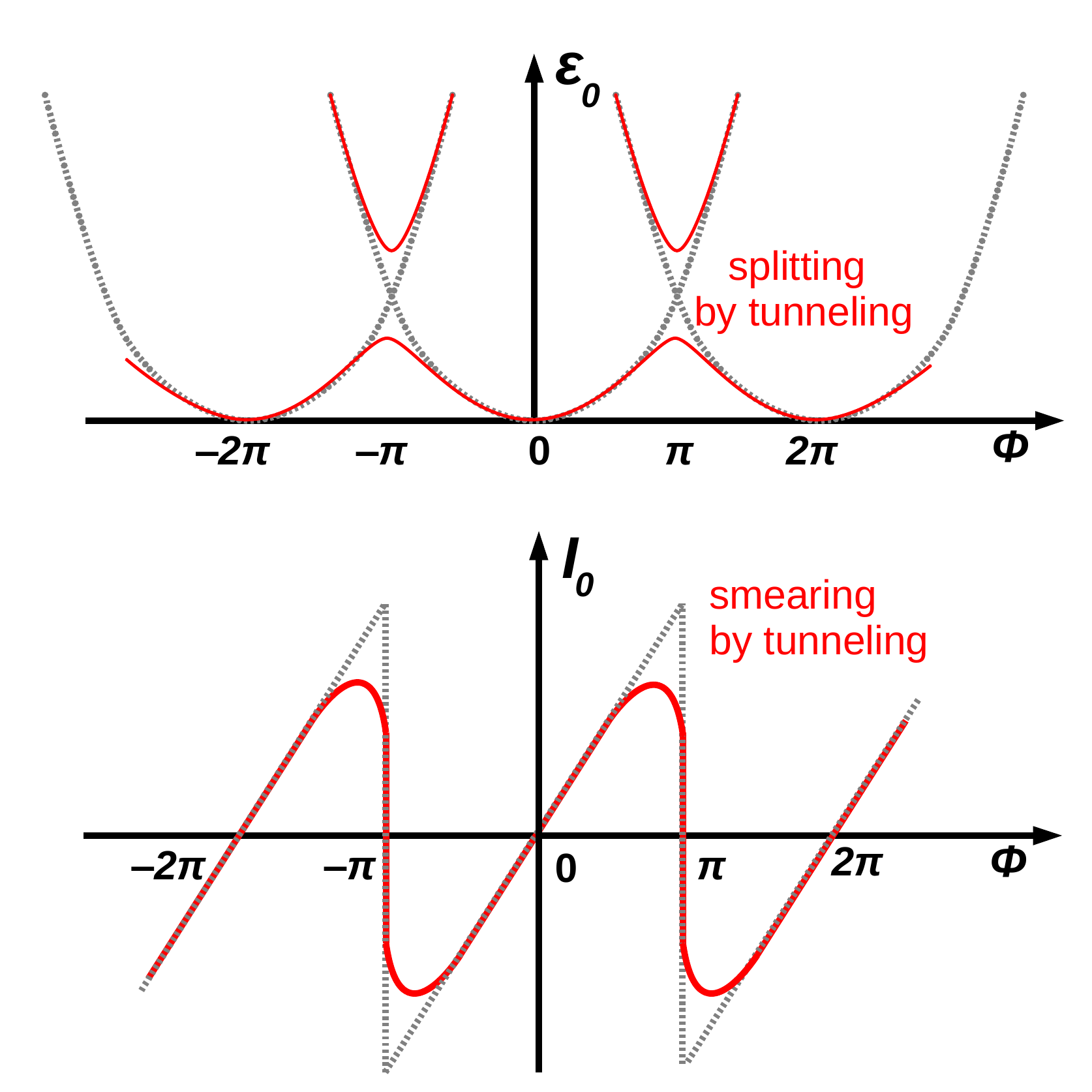}
\caption{\label{fig:Current} 
Flux dependence of the ground state energy (upper panel) and persistend current (lower panel), shown schematically in the purely classical approximation (grey dashed line) and taking into account quantum tunneling (red solid line).}
\end{figure}

The observable quantities are flux-dependent ground state energy $\mathcal{E}_0(\Phi)$, or the persistent current $I_0(\Phi)\propto\partial\mathcal{E}_0/\partial\Phi$. In the zero approximation, one can associate $\mathcal{E}_0(\Phi)$ with the static potential energy, discussed above. Then, $I_0(\Phi)$ has a discontinuous sawtooth-like dependence on~$\Phi$, as schemetically shown on Fig.~\ref{fig:Current}. Quantum tunneling results in energy splitting between the degenerate configurations when $\Phi$ is close to an odd multiple of~$\pi$, which is measurable~\cite{Astafiev2012}. Also, the sawtooth in $I_0(\Phi)$ is smoothened. A spatial modulation of the chain parameters modifies the quantum tunneling amplitude, together with the energy splitting and the smoothening of the sawtooth in $I_0(\Phi)$.

The second integration constant~$\phi_0$ cannot be found from energetic considerations, as the energy does not depend on the global phase. This does not mean, however, that $\phi_0$ can be simply dropped from the consideration. Because of the degeneracy with respect to $\phi_0$, each of the found energy minima is a circle rather than a point in the configuration space. The system eigenstates can be classified by the conjugate variable, which is the conserved total charge (the number of Cooper pairs). To estimate the tunnel splitting in the sector with zero excess charge, we can assume that the system starts from some point on the $\vartheta=\pi$ circle, which can be taken $\phi_0=0$ without loss of generality, and then sum the amplitudes of tunnelling towards different points of the $\vartheta=-\pi$ circle. For a spatially homogeneous chain, symmetry considerations fix the dominant destination at $\vartheta=-\pi$ to be $\phi_0=\pi$ \cite{Hekking1997}. In the inhomogeneous case, it should be determined by the classical trajectory, as given by Eq.~(\ref{eq:phi_0}).

\subsection{The QPS trajectory}\label{ssec:QPStrajectory}

\begin{figure}
\includegraphics[width=8cm]{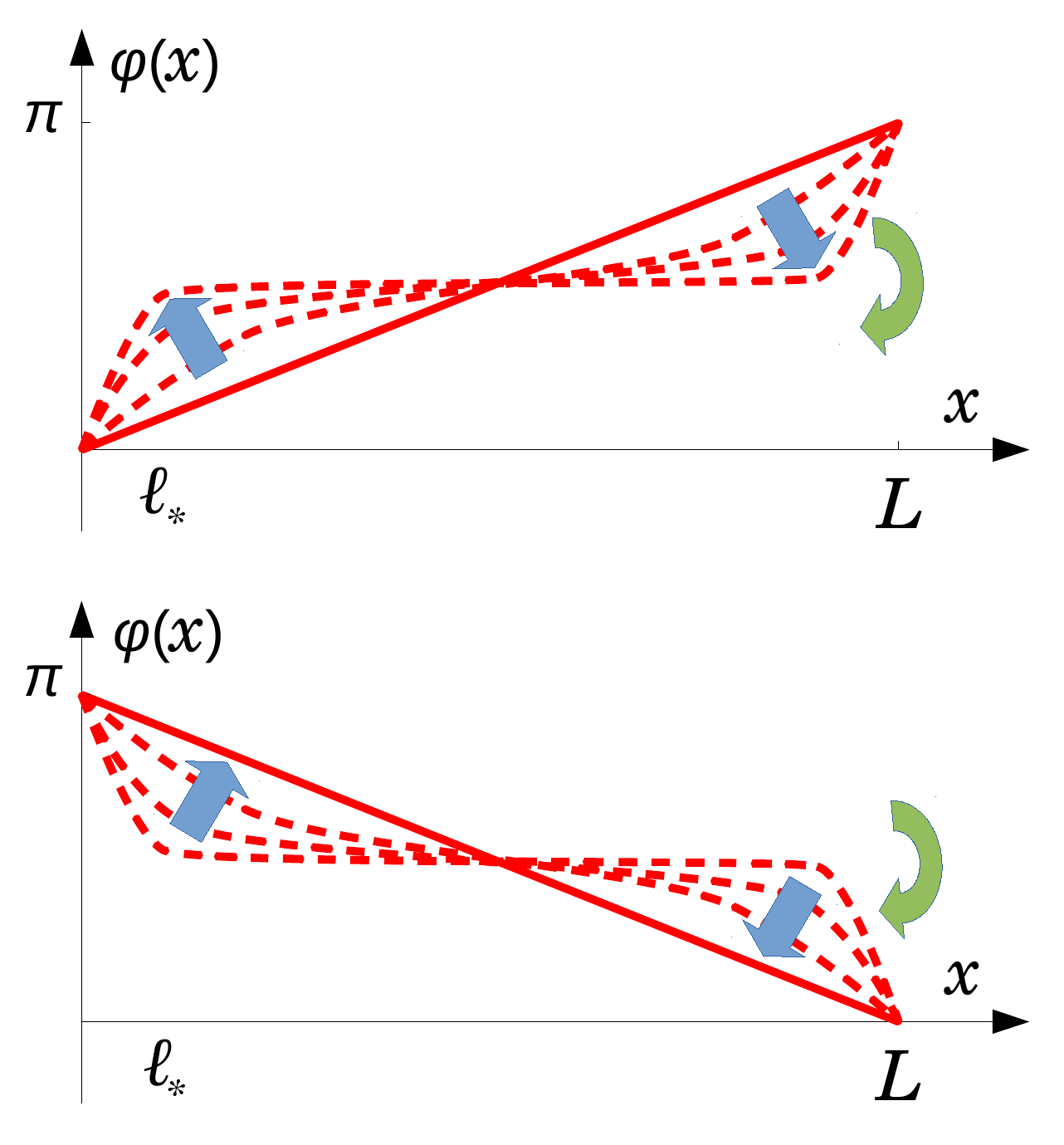}
\caption{\label{fig:QPS}
A schematic representation of the classical trajectory $\phi(x,\tau)$ going from the static configuration with $\vartheta=\pi$ (solid line on upper panel) to $\vartheta=-\pi$ (solid line on the lower panel) for a spatially homogeneous ring. Straight arrows correspond to the slow adjustment of the phase in the whole ring, the round arrows show the fast flip of the phase in the vicinity of the $\tilde{E}_J$ junction. In a ring with spatially modulated parameters, some modulation will also be superimposed on $\phi(x,\tau)$.}
\end{figure}

The main contribution to the tunnelling amplitude comes from the vicinity of the classical imaginary-time trajectory, connecting the two minima, which satisfies the Lagrange equations of motion in the imaginary time. Following the discussion of Ref.~\cite{Hekking1997} for a spatially homogeneous ring, we schematically show the corresponding configuration space trajectory $\phi(x,\tau)$ in Fig.~\ref{fig:QPS}.

The trajectory consists of several stages. (i)~Slow flattening of the phase profile in the whole chain on the time scales which are linked to the spatial scales as $\tau\sim{x}/v_\mathrm{p}$, except the vicinity of the $\tilde{E}_J$ junction. This vicinity is characterized by a certain length scale $\ell_*$ to be determined later. (ii)~Flattening of the phase in the vicinity on the time scale $\sim\ell_*/v_\mathrm{p}$. (iii)~Fast phase flip on the $\tilde{E}_J$ junction, which may occur on the same time scale $\sim\ell_*/v_\mathrm{p}$ or a faster one, depending on the parameters.  (iv),~(v)~Phase readjustment in the vicinity and outside to the new classical configuration on the same time scales as (ii) and (i), respectively. Correspondingly, the QPS action has a fast contribution from stages~(ii)--(iv) and an Ohmic environment contribution from the slow stages (i)~and~(v), dominated by the linear region of the mode dispersion.

\subsection{Summary of known results for spatially homogeneous JJ chains}
\label{ssec:summary-hom}

In Secs.~\ref{ssec:wires-junctions}--\ref{ssec:QPStrajectory} we introduced several energy and length scales. Here we summarize what is known about QPSs in spatially homogeneous JJ chains for different relations between these scales \cite{Hekking1997, Matveev2002, Vanevic2012, Rastelli2013}. For completeness, we give the calculation details in Appendix~\ref{app:homogeneous}.

For $\tilde{E}_J\lesssim{E}_J$ one can imagine two limiting cases: $\tilde{E}_J\gg\omega_\mathrm{p}$ and $\tilde{E}_J\ll\omega_\mathrm{p}$. The former case automatically applies if the $\tilde{E}_J$ junction is the same as the rest of the chain. The case $\tilde{E}_J\ll\omega_\mathrm{p}$ can be realized if $\tilde{E}_J$ is very small, $\tilde{E}_J/E_J\ll\sqrt{(2e)^2/(CE_J)}\ll{1}$. In this case, the normal modes with $\omega\gtrsim\tilde{E}_J$ remain in the ground state when a Cooper pair tunnels through the $\tilde{E}_J$ junction; the overlap between the ground states before and after the tunnelling renormalizes the tunnelling amplitude~$\tilde{E}_J$~\cite{Hekking1997, Vanevic2012}. The effect of a spatial modulation of the chain parameters on this renormalization was studied in Ref.~\cite{Taguchi2015} and will not be considered here. In the following, we assume $\omega_\mathrm{p}\lesssim\tilde{E}_J$.

The matrix element~$W$ for a QPS on the $\tilde{E}_J$ junction 
\begin{equation}
W=Ae^{-S_\mathrm{fast}-S_\mathrm{env}},
\end{equation}
where $S_\mathrm{fast}$ and $S_\mathrm{env}$ are the actions along the classical instanton trajectory corresponding to the fast motion in the vicinity of the QPS center and the Ohmic environment part, respectively, as discussed in Sec.~\ref{ssec:QPStrajectory}. The prefactor~$A$ is due to Gaussian integration over the fluctuations around the classical trajectory. These quantities have somewhat different form, depending on the relation between various length scales of the problem. Besides the chain length~$L$, assumed to be the largest scale, we have the screening length~$\ell_s$, defined in Sec.~\ref{ssec:wires-junctions}, as well as $\ell_J=e_l/\tilde{E}_J$, defined in Sec.~\ref{ssec:classicalconf}. It is convenient to introduce one more length scale,
\begin{equation}
\ell_c\equiv\frac{\ell_s}2+\frac{e_c}{\tilde{E}_c}.
\end{equation}
The Ohmic environment contribution is given by
\begin{equation}\label{eq:Shomogen}
\frac{S_\mathrm{env}}g=\ln\frac{L}{\ell_*},\quad
\ell_*\equiv\max\{4\ell_J,\ell_s+\sqrt{\ell_J\ell_c}\}.
\end{equation}
The cutoff scale $\ell_*$ determines the non-Ohmic vicinity of the $\tilde{E}_J$ junction. Eq.~(\ref{eq:Shomogen}) was obtained in Refs.~\cite{Hekking1997, Vanevic2012} and in Ref.~\cite{Rastelli2013} for the cases $\ell_*\sim\ell_J$ and $\ell_*\sim\ell_s$, respectively.

The fast non-Ohmic part of action can be evaluated in two limits:
\begin{subequations}\begin{align}
&\frac{S_\mathrm{fast}}g=0.548417 + O(\ell_c/\ell_J),\quad
\ell_J\gg\ell_c,\label{eq:Sfast}\\
&\frac{S_\mathrm{fast}}g=\frac{8}\pi
\sqrt{\frac{\ell_c}{\ell_J}}-
\Upsilon(\sqrt{\ell_J\ell_c}/\ell_s)+O(\sqrt{\ell_J/\ell_c}),
\;\;\;\ell_J\ll\ell_c,\label{eq:Score}\\
&\Upsilon(z)\equiv\int_0^\infty\left(\sqrt{1+\frac{1}{u^2}}-1\right)
\tanh^2\frac{\pi{z}u}2\,du+{}\nonumber\\
&{}\quad\qquad+1.567514-\ln(1+z),\nonumber
\end{align}\end{subequations}
where $\Upsilon(z)$ is bounded, $1.5<\Upsilon(z)<2$.
The practically important case when the $\tilde{E}_J$ junction is identical to all others, corresponds to $\ell_J\ll\ell_c=\ell_s^2/\ell_J$, $\Upsilon(1)=1.74126\ldots$.
In the mentioned limiting cases, it is also possible to evaluate the prefactor~$A$:
\begin{subequations}\begin{align}
&A=\sqrt{\frac{2\pi}g}\,\frac{e_l}{\ell_s}
\left(\frac{\zeta-1}{\zeta+1}\right)^{\zeta/4},\quad
\ell_J\gg\ell_c,\\
&\zeta\equiv\frac{e_c}{\sqrt{e_c^2-\ell_s^2\tilde{E}_c^2/4}},\nonumber\\
&A=\frac{4\tilde{E}_J}{\sqrt{g}}
\left(\frac{\ell_J}{\ell_c}\right)^{1/4},
\quad\ell_J\ll\ell_c.
\end{align}\end{subequations}
The first term in Eq.~(\ref{eq:Score}) can be rewritten as $S_\mathrm{fast}=8(\tilde{E}_J/E_c')^{1/2}$, where $E_c'=(2e)^2/(\tilde{C}+\sqrt{CC_\mathrm{g}}/2)$, similar to Refs.~\cite{Matveev2002, Rastelli2013}.
To the best of our knowledge, the numerical constant in Eq.~(\ref{eq:Sfast}) and the $\Upsilon$~term in Eq.~(\ref{eq:Score}) have not been reported in the literature before.

\subsection{Summary of our results for modulated JJ~chains}
\label{ssec:results}

We consider a JJ chain whose parameters $e_l(x)$, $e_c(x)$, $\ell_s^2(x)$ have a weak spatial modulation, for simplicity chosen to be periodic with period $L/m$. We assume $m\gg{1}$ to be integer, so that there is no discontinuity of the modulation at the $\tilde{E}_J$ junction.
We calculate the correction $\delta{S}$ to the QPS action $S_\mathrm{QPS}$ to linear order in the modulation amplitude. We consider several cases, when all three parameters are modulated,
or only one or two of them are modulated while the rest remain constant, as discussed in more detail in Sec.~\ref{ssec:modulations}.

For all these cases our main result can be expressed as
\begin{equation}\label{eq:result}
\frac{\delta{S}_\mathrm{env}}{g_0}=\frac{\delta g}{g_0}
\left[\ln\frac{L/m}{\ell_*} + O(1)\right]+O(\delta{g}^2/g_0^2).
\end{equation}
Here $g_0$ is the dimensionless admittance of the homogeneous chain, and $\delta{g}=\pi\sqrt{e_l(0)/e_c(0)}-g_0$ is the change in the local value of the admittance at the phase-slip position due to the parameter modulation.  Different modulation types (when only $e_l$ or $e_c$ is modulated, or both of them) correspond to different $\delta{g}$, and it is this $\delta{g}$ that enters Eq.~(\ref{eq:result}). The $O(1)$ term depends on the modulation type, it is calculated numerically in Sec.~\ref{sec:periodic}.  

The logarithmic term has the same form as Eq.~(\ref{eq:Shomogen}), but the logarithm is cut off at the modulation period instead of the chain length. Eq.~(\ref{eq:result}) is valid when $L/m\gg\ell_*$; otherwise, the expression in the square brackets is small. The general picture is that the QPS action is sensitive to modes whose wavelength is~$\sim\ell_*$ or larger; on the other hand, modes with the wavelength larger than modulation period are not affected by the modulation since it effectively averages out. For this reason, the long-distance cutoff of the main logarithmic term in Eq.~(\ref{eq:Shomogen}) is not modified, and the coefficient is determined by the spatial average of~$g$. This also implies that the modulation does not affect the long-distance physics of the superconductor-insulator transition in the thermodynamic limit, although the transition point may be shifted. 

We have also calculated the linear correction to the first term in Eq.~(\ref{eq:Score}),  important in the limit $\ell_J\ll\ell_c$. If the modulation period $L/m\gg\ell_s$, the correction corresponds to setting the capacitances $C$ and $C_\mathrm{g}$ to their local values at the QPS location. If the period is short, $L/m\ll\ell_s$, the modulation is averaged out and the action is determined by the spatial average of the capacitances.

\section{Spatially modulated JJ chain: general relations}
\label{sec:general}

\subsection{Change of variables and elimination of harmonic modes}
\label{ssec:variables}

Let us perform a change of variables in action~(\ref{eq:S_wire+junction}), similarly to Ref.~\cite{Rastelli2013}:
\begin{equation}
\phi(x,\tau)=\vartheta(\tau)\,X(x)
+\phi_0(\tau)+\sum_{\alpha=1}^\infty\phi_\alpha(\tau)\,\Psi_\alpha(x),
\end{equation}
where we denoted
\begin{subequations}\begin{align}
&X(x)\equiv\frac{\int_0^xe_l^{-1}(x')\,dx'}{\int_0^Le_l^{-1}(x')\,dx'}
-\frac{1}{\mathcal{T}_c}\int_0^L\frac{dx}{e_c(x)}
\frac{\int_0^xe_l^{-1}(x')\,dx'}{\int_0^Le_l^{-1}(x')\,dx'},\nonumber\\
& \label{eq:Xx=}\\ &\mathcal{T}_c\equiv\int_0^L\frac{dx}{e_c(x)},
\end{align}\end{subequations}
and $\Psi_\alpha(x)$ are the eigenfunctions of Eq.~(\ref{eq:Helmholtz}). Since $\phi_0(\tau)$ and $\vartheta(\tau)$ take care of the uniform phase shift and the phase jump between $x=0$ and $x=L$, respectively, $\Psi_\alpha(x)$ can be chosen to satisfy the Dirichlet boundary conditions, $\Psi_\alpha(0)=\Psi_\alpha(L)=0$. They are orthogonal,
\begin{equation}\label{eq:orthogonality}
\overline{(\Psi_\alpha,\Psi_\beta)}=\delta_{\alpha\beta},
\end{equation}
with the scalar product of two arbitrary functions $f_1(x)$ and $f_2(x)$ defined as
\begin{equation}
\overline{(f_1,f_2)}\equiv
\frac{1}{\mathcal{T}_c}\int_0^L\frac{dx}{e_c(x)}
\left[f_1(x)f_2(x)+\ell_s^2\,\frac{df_1(x)}{dx}\,\frac{df_2(x)}{dx}\right].
\end{equation}
The constant offset in Eq.~(\ref{eq:Xx=}) is chosen specifically to yield $\overline{(1,X)}=0$.
Using the relations
\[
\int_0^Le_l\,
\frac{d\Psi_\alpha}{dx}\frac{d\Psi_\beta}{dx}\,dx=
\mathcal{T}_c\omega_\alpha^2\delta_{\alpha\beta},\quad
\int_0^Le_l\,\frac{dX}{dx}\frac{d\Psi_\beta}{dx}\,dx=0,
\]
we rewrite action~(\ref{eq:S_wire+junction}) in the new variables (we also set $\Phi=\pi$ explicitly and add a constant for the instanton action to be finite):
\begin{align}
&\frac{S}{\mathcal{T}_c}=\int{d}\tau\left\{\frac{\dot\phi_0^2}2
+\left[\overline{(X,X)}+\frac{1}{\mathcal{T}_c\tilde{E}_C}\right]
\frac{\dot\vartheta^2}2\right.+{}\nonumber\\
&\hspace*{2cm}{}+\frac{\vartheta^2}{2\mathcal{T}_c\int_0^Le_l^{-1}(x')\,dx'}
+\frac{\tilde{E}_J}{\mathcal{T}_c}\left(1+\cos\vartheta\right)\nonumber\\
&\hspace*{2cm}{}+\sum_\alpha
\left(\frac{\dot\phi_\alpha^2}2+\frac{\omega_\alpha^2\phi_\alpha^2}2\right)
+{}\nonumber\\
&\hspace*{2cm}{}+\left.
\sum_\alpha\overline{(X,\Psi_\alpha)}\,\dot\phi_\alpha\dot\vartheta
+\sum_\alpha\overline{(1,\Psi_\alpha)}\,\dot\phi_\alpha\dot\phi_0\right\}.
\end{align}
At $L\gg\ell_s$ the inductive term, proportional to $\vartheta^2$, is negligible compared to $\tilde{E}_J\cos\vartheta$.

The classical instanton trajectory is constructed by solving the Euler-Lagrange equations. Equations for $\phi_0$ and $\phi_{\alpha>0}$ are linear, so these variables can be easily eliminated. The remaining equation for $\vartheta$ is nonlinear:
\begin{equation}\label{eq:integro-differential}
\int{K}(\tau-\tau')\,\vartheta(\tau')\,d\tau'
=\tilde{E}_J\sin\vartheta(\tau),
\end{equation}
with the kernel $K$ most easily expressed in the Fourier space:
\begin{equation}\label{eq:kernel}
K(\omega)=\omega^2\left[\frac{1}{\tilde{E}_c}+G_{XX}(\omega)
-\frac{G_{X1}(\omega)\,G_{1X}(\omega)}{G_{11}(\omega)}\right],
\end{equation}
where the Green's functions are defined as
\begin{multline}
G_{f_1f_2}(\omega)\equiv\mathcal{T}_c\left[\sum_\alpha
\frac{\omega_\alpha^2}{\omega^2+\omega_\alpha^2}\
\overline{(f_1,\Psi_\alpha)}\,\overline{(f_2,\Psi_\alpha)}\right.-\\
-\left. \sum_{\alpha}\overline{(f_1,\Psi_\alpha)}\overline{(f_2,\Psi_\alpha)}+\overline{(f_1,f_2)}\right],
\label{eq:Gf1f2=}
\end{multline}
for arbitrary $f_1(x)$, $f_2(x)$. (Note that the second line is not necessarily zero: while the functions $\Psi_\alpha(x)$ form a complete set in the space of functions with Dirichlet boundary conditions, both $1$ and $X(x)$ do not belong to this space.)
If the solution of Eq.~(\ref{eq:integro-differential}) is found, the eliminated variables $\phi_0$ and $\phi_{\alpha>0}$ can be found from
\begin{subequations}\begin{align}
&\phi_0(\omega)=-\frac{G_{1X}(\omega)}{G_{11}(\omega)}\,\vartheta(\omega),\label{eq:phi_0}\\
&\phi_\alpha(\omega)=\left[\frac{G_{1X}(\omega)}{G_{11}(\omega)}\,
\overline{(1,\Psi_\alpha)}-\overline{(X,\Psi_\alpha)}\right]
\frac{\omega^2\,\vartheta(\omega)}{\omega^2+\omega_\alpha^2}.\nonumber\\
\end{align}\end{subequations}
The same kernel $K(\tau-\tau')$ determines the action for $\vartheta$ obtained upon integration over all other modes:
\begin{align}
S[\vartheta]={}&{}\frac{1}2\int
K(\tau-\tau')\,\vartheta(\tau)\,\vartheta(\tau')\,d\tau\,d\tau'
+{}\nonumber\\
{}&{}+\int\tilde{E}_J\left[1+\cos\vartheta(\tau)\right]d\tau.
\label{eq:Sphi-1}
\end{align}
The kernel $K(\omega)$ can also be related to the chain impedance $Z(\omega)$ at complex frequencies~\cite{Vanevic2012}:
\begin{equation}
K(\omega)=\frac{\omega^2}{\tilde{E}_c}+\frac{|\omega|}{4e^2Z(i|\omega|)}.
\end{equation}

For a spatially homogeneous chain, $K(\omega)$ can be calculated exactly. Leaving the details for Appendix~\ref{app:homogeneous}, here we give its low- and high-frequency asymptotics:
\begin{equation}
K(\omega)\propto\left\{\begin{array}{ll}
|\omega|,& |\omega|\ll v_\mathrm{p}/\ell_c,\\
\omega^2,& |\omega|\gg v_\mathrm{p}/\ell_c,
\end{array}\right.
\end{equation}
where the length scale $\ell_c=\ell_s/2+e_c/\tilde{E}_c$ was introduced in Sec.~\ref{ssec:summary-hom}.
The classical trajectory $\vartheta_\mathrm{cl}(\omega)$ can be found explicitly in the two limiting cases $\ell_J\gg\ell_c$ and $\ell_J\ll\ell_c$, and the action $S_\mathrm{cl}$ along this trajectory can be evaluated (see Appendix~\ref{app:homogeneous} for details). 


\subsection{Linear response to a modulation}
\label{ssec:linearresponse}

In the following, we will assume the spatial modulation of the chain parametrs to be weak, and focus on the linear correction $\delta{S}_\mathrm{cl}$ to the classical action~$S_\mathrm{cl}$. The modulation results in a linear correction $\delta{K}(\tau-\tau')$ to the kernel for a homogeneous JJ chain, which, in turn, produces a correction $\delta\vartheta_\mathrm{cl}(\tau)$ to the classical trajectory. Note, however, that the classical trajectory was found from the condition $\delta{S}/\delta\vartheta=0$, so the correction to the action can be evaluated on the zero-approximation classical trajectory, which is most conveniently done in the Fourier space:
\begin{equation}
\delta{S}_\mathrm{cl}=\frac{1}{2}\int\frac{d\omega}{2\pi}\,
\delta{K}(\omega)\,\left|\vartheta_\mathrm{cl}(\omega)\right|^2.
\end{equation}
When calculating the correction $\delta{K}(\omega)$ to the linear order in modulations, one can ignore the last term in Eq.~(\ref{eq:kernel}). Indeed, the homogeneous chain is symmetric with respect to $x\to{L}-x$, so $1$ and $X(x)$ have different parity, and $G_{1X}(\omega)=0$. A modulation breaking this symmetry will produce $G_{1X}(\omega)$, linear in the modulation, so the last term in Eq.~(\ref{eq:kernel}) is quadratic.


First, in the limit $\ell_J\gg\ell_c$, the classical trajectory is given by (see Appendix~\ref{app:homogeneous} for details)
\begin{equation}
\vartheta_\mathrm{cl}(\omega)=\frac{2\pi}{i\omega}\,e^{-|\omega|\tau_1},
\quad\tau_1\equiv\frac{\sqrt{e_l/e_c}}{2\tilde{E}_J}.
\end{equation}
As $1/\tau_1\ll{v}_\mathrm{p}/\ell_c$, only the low-frequency asymptotics of $K(\omega)$ is needed to calculate the classical action. It is determined by the low-frequency modes which can be found from Eq.~(\ref{eq:Helmholtz}) without the second term, $\ell_s\to{0}$.

For $\ell_J\ll\ell_c$, the classical trajectory is given by
\begin{equation}
\vartheta_\mathrm{cl}(\omega)=
\frac{2\pi}{i\omega\cosh(\pi\omega\tau_2/2)},\quad
\tau_2\equiv\sqrt{\frac{\ell_c}{\tilde{E}_Je_c}},
\end{equation}
As $1/\tau_2\gg{v}_\mathrm{p}/\ell_c$, the high-frequency asymptotics of $K(\omega)$ should be taken into account. It is convenient to separate the two contributions as
\begin{equation}\label{eq:KlowKhigh=}
K(\omega)=K_\mathrm{low}(\omega)+K_2\omega^2,
\end{equation}
where $K_\mathrm{low}(\omega)$ corresponds to the first line in Eq.~(\ref{eq:Gf1f2=}) for $G_{XX}$ and remains finite at $\omega\to\infty$. In the correction to $S_\mathrm{env}$ from $K_\mathrm{low}(\omega)$, the integral converges at frequencies $\omega\sim\min\{\omega_\mathrm{p},1/\tau_2\}$. 
Thus, in both limits the correction to the logarithmic term in $S_\mathrm{env}$ can be calculated as
\begin{equation}\label{eq:deltaSloop=}
S_\mathrm{env}+\delta{S}_\mathrm{env}=
\pi^2\mathcal{T}_c\sum_\alpha\omega_\alpha\left[
\overline{(X,\Psi_\alpha)}\right]^2
\mathcal{F}_{1,2}(\omega_\alpha\tau_{1,2}),
\end{equation}
where the functions $\mathcal{F}_{1,2}(z)$ are defined as
\begin{subequations}\begin{align}
&\mathcal{F}_1(z)=\frac{2z}\pi\int_0^\infty{e}^{-2u}\,\frac{du}{z^2+u^2},\\
&\mathcal{F}_2(z)=\frac{2z}\pi\int_0^\infty\frac{1}{\cosh^2(\pi{u}/2)}
\frac{du}{z^2+u^2}.
\end{align}\end{subequations}

The coefficient $K_2$ in Eq.~(\ref{eq:KlowKhigh=}) determines the action $S_\mathrm{fast}$ for $\ell_c\gg\ell_J$; its general expression is
\begin{equation}
K_2=\frac{1}{\tilde{E}_c}
+\mathcal{T}_c\left\{\overline{(X,X)}-\sum_\alpha
\left[\overline{(X,\Psi_\alpha)}\right]^2\right\}.\label{eq:K2=}
\end{equation}
Then, $\delta{S}_\mathrm{fast}=4\,\delta{K}_2/\tau_2$.

\section{Periodically modulated JJ chain}
\label{sec:periodic}

\subsection{Physical mechanisms for the modulation}
\label{ssec:modulations}

Here we apply the general scheme, outlined in the previous section, to the simplest case of a weak periodic modulation of the chain parameters. We assume the modulation period, $L/m$, to be an integer fraction of the chain length $L$ (that is, $m\gg{1}$~is integer). This introduces no discontinuity of the JJ chain parameters at the QPS location. Thus, the modulation is assumed to have a profile
\begin{equation}
\mu(x)=1-t\cos{k}_{2m}(x-x_0),
\end{equation}
where $t\ll{1}$ is the relative modulation amplitude, $k_{2m}\equiv{2}\pi{m}/L$, and $x_0$ parametrizes the relative QPS position with respect to the modulation. One can consider different modulations, depending on their physical implementation.

When fabricating JJ chains, one can control the area of each junction. While the Josephson energy $E_J$ and the capacitance $C$ between the islands are both proportional to the junction area, the capacitance of each island to the ground is controlled by the island area.  Assuming the junction areas to be modulated and the island areas to remain constant, we arrive at the following spatial pattern of the coefficients in action~(\ref{eq:S_wire+junction}):
\begin{subequations}
\begin{equation}\label{eq:junctionmod}
e_c(x)=e_{c0},\quad
\ell_s^2(x)=\ell_{s0}^2\mu(x),\quad
e_l(x)=e_{l0}\,\mu(x).
\end{equation}
Another possible way to modulate the parameters is to vary the island areas. In this case, the ground capacitance $C_\mathrm{g}$ of each island is modulated, while $E_J$ an $C$ remain constant. This corresponds to
\begin{equation}\label{eq:islandmod}
e_c(x)=\frac{e_{c0}}{\mu(x)},\quad
\ell_s^2(x)=\frac{\ell_{s0}^2}{\mu(x)},\quad
e_l(x)=e_{l0}.
\end{equation}
Finally, each Josephson junction can be implemented as a superconducting quantum interference device (SQUID). In a magnetic field, the corresponding Josephson energy of each SQUID is sensitive to the SQUID loop area. This enables one to modulate $E_J$ independently of~$C$; this may lead to qualitatively different effects from the previous cases~\cite{Nguyen2017}. Thus, we consider the profile
\begin{equation}\label{eq:SQUIDmod}
e_c(x)=e_{c0},\quad
\ell_s^2(x)=\ell_{s0}^2,\quad
e_l(x)=e_{l0}\,\mu(x).
\end{equation}
\end{subequations}
Below we will analyze these cases separately, closely following the approach of Ref.~\cite{Taguchi2015}.

\subsection{Junction area modulation}

We start with the case of modulation~(\ref{eq:junctionmod}). First, we calculate the correction to the classical configuration:
\begin{equation}\label{eq:Xmod}
X(x)=\left(\frac{x}{L}-\frac{1}{2}\right)
+\frac{t}{k_{2m}L}\sin{k}_{2m}(x-x_0)+O(t^2).
\end{equation}
Then, we find the normal mode wave functions $\Psi_\alpha(x)$ and frequencies $\omega_\alpha$ from the modulated wave equation,
\begin{equation}
\frac\partial{\partial{x}}\,\mu(x)\,
\frac{\partial\Psi_{\alpha}}{\partial{x}}
+\kappa^2(\omega_\alpha)\,\Psi_{\alpha}=0,
\end{equation}
where $\kappa(\omega)$ denotes the inverse of the dispersion~(\ref{eq:dispersion}):
\begin{equation}
\kappa(\omega)\equiv\frac\omega{\sqrt{e_{l0}e_{c0}-\ell_{s0}^2\omega^2}}.
\end{equation}
For $t=0$ this gives the homogeneous result $\Psi_\alpha(x)=\sqrt{2/(1+k^2\ell_{s0}^2)}\sin k_{\alpha}x$ with $\kappa(\omega_\alpha)=k_\alpha=\pi\alpha/L$.

First, we use perturbation theory in $t\ll{1}$, seeking the wave function in the form
\begin{align}
\Psi_{\alpha}(x)={}&{}\sqrt{\frac{2}{1+k^2\ell_{s0}^2}}
\left(\sin k_{\alpha}x
-\frac{B_++B_-}2\cos{k}_{\alpha}x\right.+{}\nonumber\\
{}&{}+\frac{B_+}2\,\cos{k}_{\alpha+2m}x
+\frac{B_-}2\,\cos{k}_{\alpha-2m}x{}+{}\nonumber\\
{}&{}+\left.\frac{A_+}{2}\sin k_{\alpha+2m}x
+\frac{A_-}{2}\sin k_{\alpha-2m}x\right).
\label{eq:Psialpha=}
\end{align}
The perturbation theory gives
\begin{subequations}\begin{align}
&A_\pm=-\frac{k_\alpha k_{\alpha\pm{2}m}}%
{k_\alpha^2-k_{\alpha\pm{2m}}^2}\,t\cos k_{2m}x_0,\\
&B_\pm=\pm\frac{k_\alpha k_{\alpha\pm{2}m}}%
{k_\alpha^2-k_{\alpha\pm{2}m}^2}\,t\sin k_{2m}x_{0},
\end{align}\label{eq:ABpm=}\end{subequations}
and the correction to $\omega_\alpha$ is $O(t^2)$.
\begin{widetext}\begin{align}\label{eq:XPsia}
\overline{(X,\Psi_\alpha)}
={}&{}-\frac{1+(-1)^\alpha}2\sqrt{\frac{2}{1+k_\alpha^2\ell_{s0}^2}}\,
\frac{1}{k_\alpha{L}}\left[1-\frac{t}4\cos{k}_{2m}x_0
\left(\frac{k_\alpha^2}{k_\alpha^2-k_m^2}
+2\delta_{\alpha,2m}\right)+O(t^2)\right]+\nonumber\\
&{}-\frac{1-(-1)^\alpha}2\sqrt{\frac{2}{1+k_\alpha^2\ell_{s0}^2}}\,
\frac{2t\sin{k}_{2m}x_0}{k_\alpha{k}_{2m}L^2}+O(t^2).
\end{align}\end{widetext}
However, the perturbative expression~(\ref{eq:Psialpha=}) is not always valid. By a direct check, we see that the corrections are small when two conditions are fulfilled:
\begin{equation}\label{eq:PTvalidity}
|\alpha-m|\gg{t}m,\quad t\alpha\ll{m}.
\end{equation}
The first condition breaks down in the relatively narrow interval of $\alpha$, where the gap in the frequency spectrum opens up. The resulting modification of a relatively small number of terms in the $\alpha$ sum
in Eq.~(\ref{eq:deltaSloop=}), those with $|\alpha-m|\sim{t}m$, leads to a small correction to the $L/m$ factor inside the logarithm in Eq.~(\ref{eq:result}). This correction is beyond our precision.

For large $\alpha$, the second condition~(\ref{eq:PTvalidity}) breaks down. Then, instead of doing perturbation theory, one can construct $\Psi_\alpha(x)$ using the WKB approximation:
\begin{subequations}\begin{align}\label{eq:PsiWKB=}
&\Psi_\alpha(x)=\sqrt{\frac{2}{1+\ell_{s0}^2\kappa^2(\omega_\alpha)}}\,
\frac{\sin{s}(x)}{[\mu(x)]^{1/4}},\\
&s(x)\equiv\int_0^x\frac{\kappa(\omega_\alpha)\,dx'}%
{\sqrt{\mu(x')}}.\label{eq:WKBphase}
\end{align}\end{subequations}
The frequency $\omega_\alpha$ is determined by the boundary condition for $\Psi_\alpha(x)$, that is, $s(L)=\pi\alpha$. This results in a small relative correction $O(t^2)$ to the frequency and determines the normalization factor in Eq.~(\ref{eq:PsiWKB=}). Although the relative difference between $s(x)$ and its zero-approximation value $k_\alpha{x}$ is small, the absolute difference may become of the order of one, and then $\sin{s}(x)-\sin{k}_\alpha{x}\sim{1}$ as well. This is the reason of the perturbation theory breakdown at large~$\alpha$. Note, however, that the perturbation theory is valid at $\alpha\ll{m}/t$, while the WKB approximation is valid at $\alpha\gg{m}$, so their regions of validity overlap.

\begin{figure}
\includegraphics[scale=0.5]{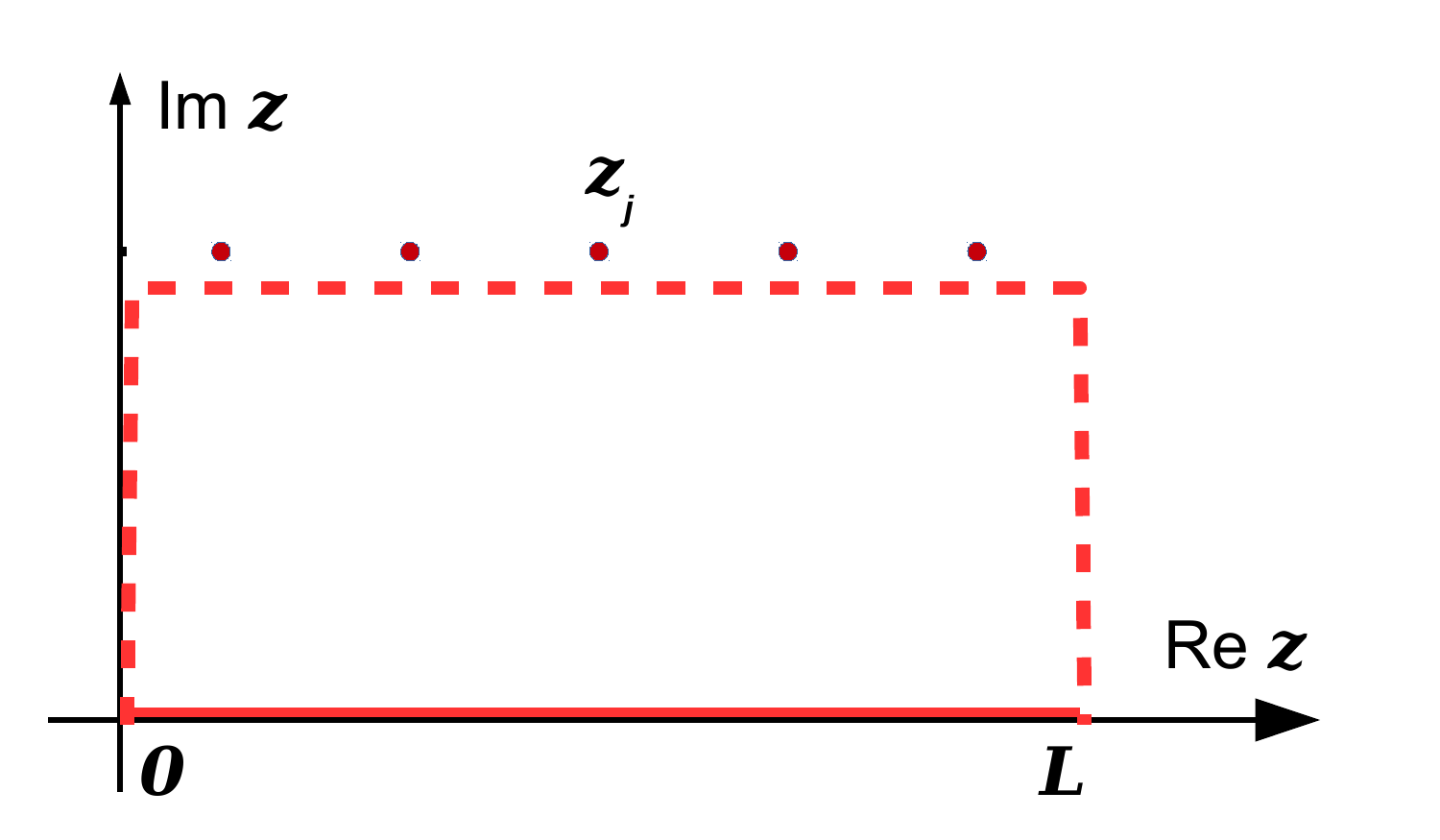}
\caption{\label{fig:contour} (Color online)
Deformation of the integration contour in Eq.~(\ref{eq:expis}) from the real axis (solid red line) into the upper complex half-plane (dashed red line). The dots represent branching points $z_j$ of $s(z)$.}
\end{figure}

Now we evaluate the overlap $\overline{(X,\Psi_{\alpha})}$ writing it as
\begin{align}
\overline{(X,\Psi_{\alpha})}={}&{}
\sqrt{\frac{2}{1+k_\alpha^2\ell_{s0}^2}}\Im\int_0^L
\frac{dx}{L}\,e^{is(x)}[\mu(x)]^{1/4}\times{}\nonumber\\
{}&{}\times\left[\frac{X(x)}{\sqrt{\mu(x)}}
+ik_\alpha\ell_{s0}^2\,\frac{dX(x)}{dx}\right].
\label{eq:expis}
\end{align}
Note that $e^{is(x)}$ is fast oscillating, while the rest of the integrand is smooth, due to the condition $k_\alpha\gg{k}_{2m}$. Thus, we introduce the complex variable $z$ such that $x=\Re{z}$, and deform the contour into the upper complex half-plane, as shown in Fig.~\ref{fig:contour}. The contour can be moved up to the branching points of $s(z)$, located at 
\[
z_j=\frac{jL}m+x_0+\frac{i}{k_{2m}}\arccosh\frac{1}t.
\]
The integral over the horizontal part of the contour near the branching points is suppressed as $t^{\alpha/(2m)}$; the branching points determine the small reflection probability from a weak smooth potential, which in the present case of a periodic modulation leads to opening of small gaps at high frequencies. This effect is beyond our precision, so the contribution of interest comes from the steepest descent in the positive imaginary direction from the points $x=0$ and $x=L$. To linear order in~$1/k_\alpha$ this gives
\begin{equation}
\overline{(X,\Psi_{\alpha})}=
\sqrt{\frac{2}{1+k_\alpha^2\ell_{s0}^2}}\,
\frac{[\mu(0)]^{1/4}}{k_\alpha{L}}
\left[X(0)-(-1)^\alpha{X}(L)\right].
\end{equation}
This coincides with Eq.~(\ref{eq:XPsia}) in the limit $\alpha\gg m$. The reason for this coincidence is that even though the WKB wave function differs significantly from the perturbative one in the bulk of the chain, the overlap integral is dominated by the vicinities of $x=0,L$, where the phase accumulated in $s(x)$ is still small on the absolute scale.

Thus, Eq.~(\ref{eq:XPsia}) can be used for all~$\alpha$.
Substituting it into Eq.~(\ref{eq:deltaSloop=}) and neglecting $O(t^2)$ terms, we obtain
\begin{equation}
\frac{\delta{S}_\mathrm{env}}g=-\frac{\pi{t}}{L}\cos{k}_{2m}x_0
\sum_{\alpha\:\mathrm{even}}\frac{k_\alpha}{k_\alpha^2-k_m^2}\,
\frac{\mathcal{F}_{1,2}(\omega_\alpha\tau_{1,2})}%
{(1+k_\alpha^2\ell_{s0}^2)^{3/2}}.
\end{equation}
The sum can be replaced by the integral which should be understood as the principal value (the contribution of the term with $\alpha=2m$ has relative smallness $\sim{1/m}$). The last factor cuts off the integral at $k_\alpha\sim{1}/\ell_*$. At $k_m\ell_*\ll{1}$ the integral is logarithmic, where the small~$k$ cutoff is determined by the first factor. In this case it is convenient to rewrite it as
\begin{equation}
\frac{\delta{S}_\mathrm{env}}g=-\frac{t\cos{k}_{2m}x_0}2\int_0^\infty\frac{\mathcal{F}_{1,2}(\omega(k)\tau_{1,2})\,dk}%
{(k+k_m)(k^2\ell_{s0}^2+1)^{3/2}},
\end{equation}
where we used the fact that the integral of $k/(k^2-k_m^2)-1/(k+k_m)$ is identically zero. As a result,
\begin{equation}
\frac{\delta{S}_\mathrm{env}}g=-\frac{t}2\cos{k}_{2m}x_0
\left(\ln\frac{1}{k_m\ell_*}+\tilde\Upsilon\right),\label{eq:deltaSenv}
\end{equation}
if $k_m\ell_*\ll{1}$; at $k_m\ell_*\gg{1}$, the correction is suppressed as $\sim{1}/(k_m\ell_*)$ or $1/(k_m\ell_*)^2$, depending on the limiting case.
$\tilde\Upsilon$ is a number of the order of unity, evaluated numerically. In the limit $\ell_J\gg\ell_c$, we obtain $\tilde\Upsilon=\ln{4}-\gamma\approx{0}.809\ldots$ within our numerical precision ($\gamma=0.577\ldots$ is the Euler-Mascheroni constant).
For $\ell_J\ll\ell_c$, $\tilde\Upsilon$ becomes a function of $\sqrt{\ell_J\ell_c}/\ell_s$, which varies in a finite interval: at $\sqrt{\ell_J\ell_c}/\ell_s\to{0}$, we obtain analytically $\tilde\Upsilon=\ln{2}-1=-0.3069\ldots$, while at $\sqrt{\ell_J\ell_c}/\ell_s\to\infty$, $\tilde\Upsilon=-0.6611\ldots$. At $\sqrt{\ell_J\ell_c}/\ell_s=1$, it amounts to $\tilde\Upsilon=-0.4806\ldots$.
For realistic parameters, e.~g., a chain of 1000 junctions with $g=3$ and $\ell_s=\sqrt{\ell_c\ell_J}=10\,a$, modulated with $t=0.2$ and $m=5$, this gives $\delta{S}_\mathrm{env}\approx{0}.2$.


Finally, to find the correction to the high-frequency asymptotics of the kernel $K(\omega)$, determined by Eq.~(\ref{eq:K2=}), we directly evaluate
\begin{align}\label{eq:deltaK2}
&\overline{(X,X)}-\sum_\alpha\left[\overline{(X,\Psi_\alpha)}\right]^2={}
\nonumber\\
{}&{}=\frac{\ell_{s0}}{2L}\left(1
-\frac{t}2\,\frac{\cos{k}_{2m}x_0}{1+k_m^2\ell_{s0}^2}\right)
+O(\ell_s^2/L^2).
\end{align}
For $k_m\ell_{s0}\ll{1}$, this correction corresponds precisely to the local value of $\ell_s$, and thus of $\ell_s^2/e_c\propto{C}$ at the QPS location. For $k_m\ell_{s0}\gg{1}$, the correction is suppressed, as the modulation is effectively averaged out on the length~$\ell_{s0}$, as discussed in Sec.~\ref{ssec:results}.

\subsection{Island area modulation}

For modulation~(\ref{eq:islandmod}), Eq.~(\ref{eq:Xx=}) gives 
\begin{align}
X(x)={}&{}\frac{x}{L}-\frac{1}{2}
-\frac{t}{k_{2m}L}\sin{k}_{2m}x_0.
\end{align}
The wave functions $\Psi_\alpha$ are found from the wave equation
\begin{equation}
\frac{\partial^2\Psi_\alpha}{\partial{x}^2}
+\kappa^2(\omega_\alpha)\,
\mu(x)\,\Psi_\alpha=0.
\end{equation}
The perturbative expression for $\Psi_\alpha(x)$ is again Eq.~(\ref{eq:Psialpha=}), with coefficients obtained from Eqs.~(\ref{eq:ABpm=}) by replacing $k_{\alpha\pm{2}m}\to{k}_\alpha$ in the numerators and inverting the overall sign. The WKB wave function is given by the same expression~(\ref{eq:PsiWKB=}), but instead of Eq.~(\ref{eq:WKBphase}), the phase $s(x)$ is given by
\begin{equation}
s(x)=\kappa(\omega_\alpha)\int_0^x
\sqrt{\mu(x')}\,dx'.
\end{equation}
The final result for $\overline{(X,\Psi_{\alpha})}$ turns out to be exactly the same as for the case of the junction area modulation, Eq.~(\ref{eq:XPsia}). $\delta{S}_\mathrm{env}$ is also given by Eq.~(\ref{eq:deltaSenv}).

Evaluation of Eq.~(\ref{eq:K2=}) with the perturbed wave functions again gives Eq.~(\ref{eq:deltaK2}). This time, at $k_m\ell_{s0}\ll{1}$ it corresponds to taking the local value of the ground capacitance~$C_\mathrm{g}$.

\subsection{SQUID area modulation}

For modulation~(\ref{eq:SQUIDmod}), the profile $X(x)$ is again given by Eq.~(\ref{eq:Xmod}). The coefficients $A_\pm,B_\pm$ are obtained by multiplying those from Eqs.~(\ref{eq:ABpm=}) by $1+k_\alpha^2\ell_s^2$. All subsequent calculations are analogous; the result is the same as in Eq.~(\ref{eq:deltaSenv}) but the number $\tilde\Upsilon$ is different in the limit $\ell_J\ll\ell_c$. At $\sqrt{\ell_J\ell_c}/\ell_s\to{0}$, we have $\tilde\Upsilon=\ln{2}$, at $\sqrt{\ell_J\ell_c}/\ell_s\to\infty$, $\tilde\Upsilon=-0.6611\ldots$, and at $\sqrt{\ell_J\ell_c}=\ell_s$ we obtain $\tilde\Upsilon=-0.0695\ldots$.

Evaluation of Eq.~(\ref{eq:K2=}) can be simplified by noting that modulation~(\ref{eq:SQUIDmod}) does not affect the scalar product. By completeness, $\sum_\alpha\Psi_\alpha(x)\Psi_\alpha(x')\equiv\mathcal{I}(x,x')$ is the kernel of the unit operator in the space of functions with Dirichlet boundary conditions, and it does not depend on the choice of the functional basis~$\Psi_\alpha$ in this space. Thus, Eq.~(\ref{eq:K2=}) can be evaluated using the wave functions for the homogeneous chain, $\Psi_\alpha(x)=\sqrt{2/(1+k_\alpha^2\ell_{s0}^2)}\sin k_{\alpha}x$. As a result, the correction vanishes. Indeed, modulation~(\ref{eq:SQUIDmod}) does not involve the capacitances at all.

\subsection{Combined modulation}
\label{ssec:superposition}

We can also consider a case when both Josephson energies and capacitances are modulated, $e_l(x)=e_{l0}\,\mu_l(x)$ and $e_c(x)=e_{c0}/\mu_c(x)$, generally speaking, with two different amplitudes $t_l$~and~$t_c$. Then, it is easy to see that the resulting effect on $\overline{(X,\Psi_{\alpha})}$ is additive. For the first-order perturbative wave functions this follows trivially, while for the WKB wave functions it follows from the steepest-descent calculation, analogous to Eq.~(\ref{eq:expis}). Its result is determined by the derivative $s'(x=0)$, which, in turn, can be calculated perturbatively.

The results obtained above may be conveniently combined if we introduce the local dimensionless admittance:
\begin{equation}\label{eq:admittance}
g(x)\equiv\pi\sqrt{\frac{e_l(x)}{e_c(x)}}\equiv
g_0+\delta g(x).
\end{equation}
For all types of modulation, discussed in Sec.~\ref{ssec:modulations}, we have $\delta{g}(x)/g_0=-(t/2)\cos k_{2m}\left(x-x_0\right)+O(t^2)$. For the combined modulation with two different amplitudess $t_l$ and $t_c$, the correction is $\delta{g}(x)/g_0=-(t_l/2+t_c/2)\cos k_{2m}\left(x-x_0\right)+O(t^2)$. Then, up to terms $O(1)$, at $k_m\ell_*\ll{1}$ we can express correction $\delta{S}_\mathrm{env}$ in terms of $\delta{g}(x=0)$ for all types of modulations, matching Eq.~(\ref{eq:result}):
\begin{equation}
\delta S_\mathrm{env}=\delta{g}(0)\,\ln\frac{1}{k_m\ell_*},
\end{equation}


\section{Modulated superconducting wires}
\label{sec:wires}

We finish our study by discussing applicability of our results, derived for JJ chains, to the case of thin superconducting wires. The Mooij-Sch\"on modes with low frequencies, which determine the Ohmic environment, are quite similar in the two cases. The difference is that while in the JJ chain model there are no excitations above the cutoff frequency~$\omega_\mathrm{p}$, in a wire the role of the cutoff frequency is played by the superconducting gap~$2\Delta$, above which quasiparticle excitations are present and can be virtually excited during the phase tunnelling process. Thus, for the Ohmic part of the action one can use the expressions derived in this paper if $\ell_s$ is defined as the inverse cutoff wave vector: $\ell_s\sim\sqrt{e_le_c}/\Delta$. Moreover, in the limit $\ell_J\gg\ell_c$, the non-Ohmic part should also be equivalent for chains and wires, since the instanton duration $\tau_1$ is longer than the inverse cutoff frequency, so the high-energy excitations do not matter~\cite{Hekking1997, Vanevic2012}.

The non-Ohmic contribution to the action is significantly different for wires and JJ chains when $\ell_J\ll\ell_c$. A quantitative theory for the non-Ohmic contribution to the action $S_\mathrm{fast}$ in superconducting wires still does not exist. Still, some qualitative understanding can be reached. The key fact is that for wires, the instanton duration in the limit $\ell_J\ll\ell_c$ is of the order of $\Delta^{-1}$~\cite{Vanevic2012}. Then, the contribution to the action from the integral of $K(\omega)$ is parametrically smaller than that from the Josphson $\tilde{E}_J$ term (for wires the Josephson term has a more complicated form, non-local in time, but the corresponding contributions can still be identified and estimated~\cite{Vanevic2012}). Thus, $S_\mathrm{fast}$ is determined not by the length $\ell_c$, but by the superconducting coherence length~$\xi$, the shortest length scale in the theory. As a result, modulations with $L/m\ll\ell_*$ are not averaged out and $S_\mathrm{fast}$ is determined by the local values of wire parameters at the QPS position. Only very short-wavelength modulations with period $L/m\ll\xi$ average out.

\section{Conclusions and outlook}
\label{sec:conclusions}

We have analyzed coherent QPSs in a superconducting Josephson junction ring, whose parameters are subject to a weak periodic modulation in space. We calculated the correction to the QPS semiclassical action, linear in the modulation strength. We have shown that this correction is large when the modulation period is larger than the size of the non-Ohmic vicinity of the junction on which the QPS occurs; in that case, it is determined by the local value of the chain admittantce at the phase-slip position and by a logarithmic factor whose long-distance cutoff is the modulation period, in contrast to the main term in the action where the cutoff is the system length.

Our results can be extended to other spatial profiles of the modulation. Indeed, the superposition principle for the first-order correction discussed in Sec.~\ref{ssec:superposition} remains valid for a combination of modulations with different periods. An arbitrary modulation can be expanded in the Fourier series, and the effects of different terms can be added up. Thus, arbitrary modulations can be described, as long as they are relatively weak. In particular, random inhomogeneities will be studied in the future.

\acknowledgements

We thank O. Buisson and G. Rastelli for illuminating discussions. This work was supported by the French Agence Nationale de la Recherche (ANR) under grant ANR-15-CE30-0021 ``QPSNanoWires''.


\appendix

\section{QPS in a spatially homogeneous JJ chain}
\label{app:homogeneous}

\subsection{Kernel $K(\omega)$}

For a spatially homogeneous chain we have
\begin{eqnarray*}
&&\Psi_\alpha(x)=\sqrt{\frac{2}{1+k_\alpha^2\ell_s^2}}\sin{k}_\alpha{x},\quad
k_\alpha=\frac{\pi\alpha}L,\\
&&X(x)=\frac{x}L-\frac{1}2,\;\;\;
\overline{(X,\Psi_\alpha)}=
-\sqrt{\frac{2}{1+k_\alpha^2\ell_s^2}}\,\frac{1+(-1)^\alpha}{2\pi\alpha},\\
&&\overline{(X,X)}-\sum_\alpha\overline{(X,\Psi_\alpha)}^2
=\sum_{n=-\infty}^\infty\frac{\ell_s^2/L^2}{1+(2\pi{n}\ell_s/L)^2}={}\\
&&\qquad{}=\frac{\ell_s}{2L}\coth\frac{L}{2\ell_s}.
\end{eqnarray*}
The kernel $K(\omega)$ can be calculated exactly by evaluating the sum over $\alpha$ in Eq.~(\ref{eq:Gf1f2=}) for $G_{XX}$ ($G_{1X}$ vanishes by parity):
\begin{equation}\label{eq:Kexact=}
K(\omega)=\frac{\omega^2}{\tilde{E}_c}+
\frac{\ell_s\omega^2}{2e_c}\sqrt{1+\frac{\omega_\mathrm{p}^2}{\omega^2}}
\coth\frac{L}{2\ell_s}\sqrt{\frac{\omega^2}{\omega^2+\omega_\mathrm{p}^2}}
-\frac{e_l}L.
\end{equation}
We will mostly work with the $L\to\infty$ limit of this expression~\cite{Korshunov1989}
\begin{equation}
K(\omega)=\frac{\omega^2}{\tilde{E}_c}+
\frac{\ell_s\omega^2}{2e_c}\sqrt{1+\frac{\omega_\mathrm{p}^2}{\omega^2}},
\end{equation}
whose low- and high-frequency asymptotics are
\begin{subequations}\begin{eqnarray}
&&K(\omega\ll\omega_K)=\sqrt{\frac{e_l}{e_c}}\,\frac{|\omega|}2,
\label{eq:Klowfreq}\\
&&K(\omega\gg\omega_K)=
\left(\frac{1}{\tilde{E}_c}+\frac{\ell_s}{2e_c}\right)\omega^2,
\label{eq:Khighfreq}\\
&&\omega_K\equiv\frac{\sqrt{e_le_c}}{e_c/\tilde{E}_c+\ell_s/2}.
\end{eqnarray}\end{subequations}

\subsection{Classical trajectory}

First, let us study the case
\begin{equation}\label{eq:lJgg}
\ell_J\equiv\frac{e_l}{\tilde{E}_J}\gg
\frac{e_c}{\tilde{E}_c}+\frac{\ell_s}2\equiv\ell_c.
\end{equation}
It can be checked directly that the function
\begin{subequations}
\begin{equation}\label{eq:arctan}
\vartheta(\tau)=-2\arctan\frac\tau{\tau_1},\quad
\tau_1\equiv\frac{\sqrt{e_l/e_c}}{2\tilde{E}_J},
\end{equation}
with the Fourier tranform
\begin{equation}\label{eq:omegaexpomega}
\vartheta(\omega)=\frac{2\pi}{i\omega}\,e^{-|\omega|\tau_1},
\end{equation}\end{subequations}
satisfies Eq.~(\ref{eq:integro-differential}) with the kernel~(\ref{eq:Klowfreq}). This approximation is consistent because condition~(\ref{eq:lJgg}) ensures that $1/\tau_1\ll\omega_K$. Then, the instanton action is given by
\begin{equation}
S_\mathrm{cl}=\frac{1}{2}\int\frac{d\omega}{2\pi}\,K(\omega)
\left|\vartheta(\omega)\right|^2
+\int\tilde{E}_J\left[1+\cos\vartheta(\tau)\right]d\tau.
\end{equation}
The last term equals $\pi\sqrt{e_l/e_c}\equiv{g}$, while in the first term the integral is logarithmically divergent at $\omega\to{0}$. To handle this divergency, one has to go back to Eq.~(\ref{eq:Kexact=}).
At $\omega\ll\omega_K$ this amounts to replacing Eq.~(\ref{eq:Klowfreq}) by
\begin{equation}
K(\omega\ll\omega_K)=\frac{e_l}{L}
\left[\frac{L}{2\ell_s}\frac{\omega}{\omega_\mathrm{p}}
\coth\left(\frac{L}{2\ell_s}\frac{\omega}{\omega_\mathrm{p}}\right)-1\right].
\end{equation}
Strictly speaking, the solution is no longer given by Eq.~(\ref{eq:omegaexpomega}); however, the $1/\omega$ behavior at $\omega\to{0}$ is unchanged since it is determined by the overall change of $\vartheta(t)$ from $t\to-\infty$ to $t\to\infty$.
The resulting action is given by
\begin{eqnarray}
&&\frac{S_\mathrm{cl}}g=1+\int_0^\infty\frac{du}{u^2}\,(u\coth{u}-1)\,
e^{-(2\ell_J/L)u}={}\nonumber\\
&&\quad{}=\ln\frac{L}{\ell_J}+c_1+O(\ell_J/L),
\end{eqnarray}
where the constant $c_1=-0.837877\ldots$ is easily calculated numerically.

In the opposite limit, $\ell_J\ll\ell_c$, we start with the function
\begin{equation}\label{eq:arctansinh}
\vartheta(\tau)=-2\arctan\sinh\frac{\tau}{\tau_2},\quad
\tau_2\equiv\sqrt{\frac{1}{\tilde{E}_J}
\left(\frac{1}{\tilde{E}_c}+\frac{\ell_s}{2e_c}\right)},
\end{equation}
whose Fourier transform is
\begin{equation}\label{eq:omegacoshomega}
\vartheta(\omega)=\frac{2\pi}{i\omega\cosh(\pi\omega\tau_2/2)}.
\end{equation}
This function is the exact solution of Eq.~(\ref{eq:integro-differential}) with the kernel~(\ref{eq:Khighfreq}), which then describes a usual pendulum. The condition $\ell_J\ll\ell_c$ ensures that $1/\tau_2\gg\omega_K$, so expression (\ref{eq:omegacoshomega}) is valid everywhere except the narrow frequency range $|\omega|\lesssim\omega_K$. Indeed, the low-frequency expansion of Eq.~(\ref{eq:omegacoshomega}) is
\begin{equation}
\vartheta(\omega)=\frac{2\pi}{i\omega}\left[
1-\frac{\pi^2}{8}\,\omega^2\tau_2^2+O(\omega^4\tau_2^4)\right],
\end{equation}
while the analogous expansion of solution~(\ref{eq:omegaexpomega}) contains a term proportional to $|\omega|$ in the square brackets. The solution is expected to have the same analytical properties in both limiting cases, so we have to study the low-frequency region in more detail. Indeed, the presence of the $|\omega|$ term indicates that the trajectory $\vartheta(\tau)$ very slowly reaches its limiting values $\pm\pi$, which is due to coupling with the slow Ohmic modes of the chain.

\begin{figure}
\includegraphics[width=8cm]{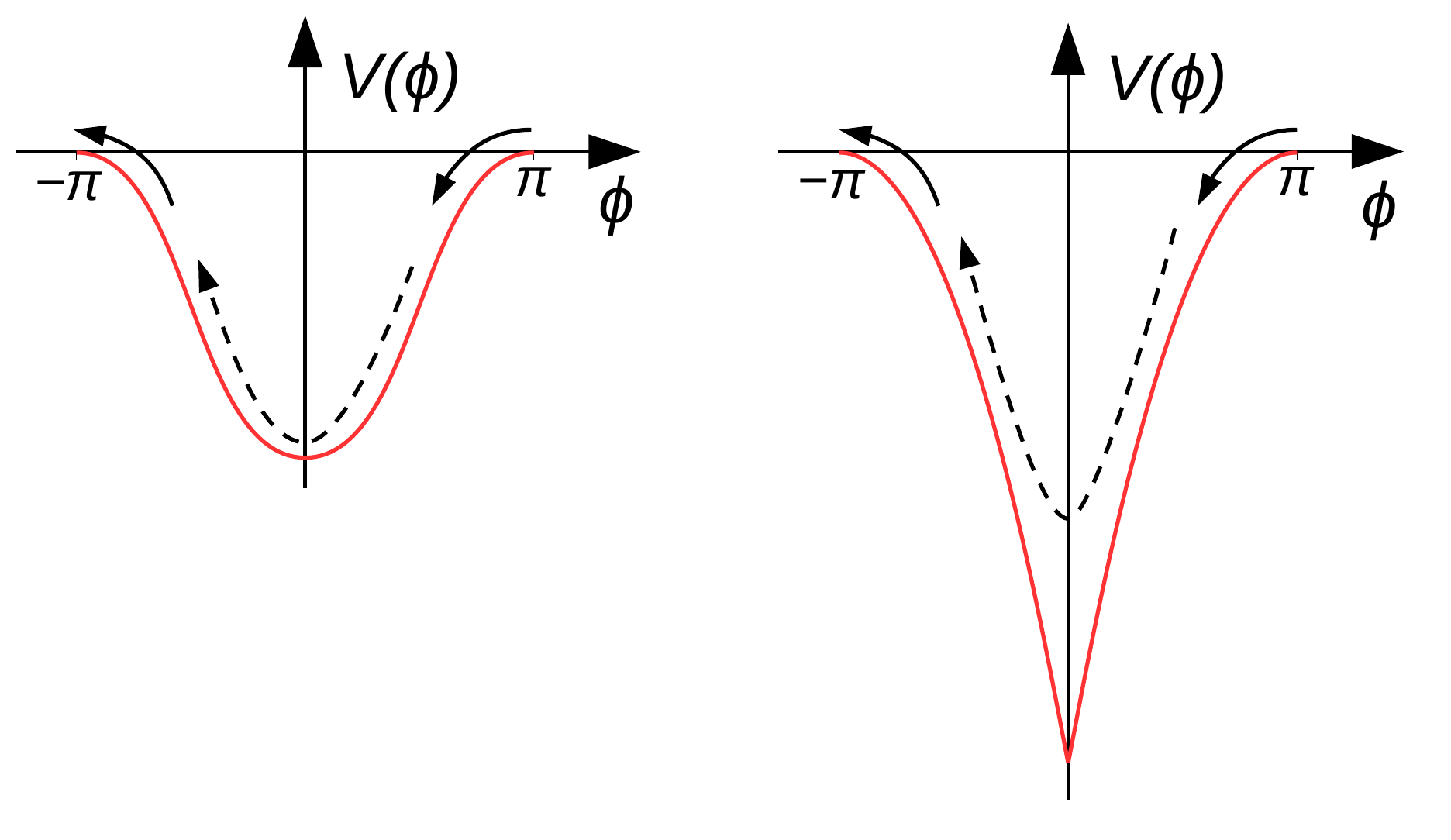}
\caption{\label{fig:potential}
The two potentials, $V(\phi)=-\tilde{E}_J(1+\cos\phi)$ (left panel) and $V(\phi)=-\tilde{E}_J(|\phi|-\pi)^2/2$ (right panel), for which the slow part of the instanton trajectory (solid arrows) should be similar.
}
\end{figure}

To analyze the slow part of the trajectory $\vartheta(\tau)$, we note that it is mostly determined by the motion near the maxima of the potential at $\vartheta=\pm\pi$. Thus, if one replaces the potential
\[
V(\phi)=-\tilde{E}_J(1+\cos\phi)\;\;\;\to\;\;\;V(\phi)=-\tilde{E}_J\,
\frac{(|\phi|-\pi)^2}2,
\]
the low-frequency part of the trajectory at $|\omega|\ll{1}/\tau_2$ should remain similar. Then, instead of Eq.~(\ref{eq:integro-differential}) we have
\begin{equation}\label{eq:Villain}
\int{K}(\tau-\tau')\,\vartheta(\tau')\,d\tau'
=\tilde{E}_J\left[\pi-|\vartheta(\tau)|\right]\sign\vartheta(\tau).
\end{equation}
This is still a non-linear equation. However, if one introduces a new variable
\begin{equation}
\tilde\vartheta(\tau)=\vartheta(\tau)+\pi\sign\tau,
\end{equation}
and uses the fact that $\sign\vartheta(\tau)=-\sign\tau$, it is easy to see that $\vartheta\phi$ satisfies a linear equation which is most easily written in the Fourier space
\begin{equation}
K(\omega)\left[\tilde\vartheta(\omega)+\frac{2\pi}{i\omega}\right]
=-\tilde{E}_J\tilde\vartheta(\omega),
\end{equation}
and gives
\begin{equation}\label{eq:KEJ}
\vartheta(\omega)=\frac{2\pi}{i\omega}\,\frac{1}{1+K(\omega)/\tilde{E}_J}.
\end{equation}
This expression has the required $|\omega|$ term at low frequencies, but
if one expands this expression in the powers of $K/\tilde{E}_J\sim\omega^2\tau_2^2\ll{1}$, the $\omega^2\tau_2^2$ term already does not match the expansion of $\cosh(\pi\omega\tau_2/2)$. However, Eq.~(\ref{eq:KEJ}) shows that the relative error of the expression~(\ref{eq:omegacoshomega}) is $\sim|\omega|\omega_\mathrm{p}\ell_s/(e_c\tilde{E}_J)$, so the relative error in the action evaluated on the trajectory~(\ref{eq:omegacoshomega}) will be of the order of $\omega_\mathrm{p}\ell_s/(e_c\tilde{E}_J\tau_2)\sim\sqrt{\ell_J/\ell_c}$.

To evaluate the action on the trajectory~(\ref{eq:omegacoshomega}), we represent $1/\cosh^2=1-\tanh^2$ and notice that in the term with $\tanh^2$ the limit $L\to\infty$ can be taken directly:
\begin{widetext}
\begin{align*}
&S_\mathrm{cl}=\int\limits_{-\infty}^{\infty}
\left[\frac{1}{\tilde{E}_c}
+\frac{\ell_s}{2e_c}\sqrt{1+\frac{\omega_\mathrm{p}^2}{\omega^2}}
\coth\frac{L}{2\ell_s}\sqrt{\frac{\omega^2}{\omega^2+\omega_\mathrm{p}^2}}
-\frac{e_l}{L\omega^2}\right]\frac{\pi\,d\omega}{\cosh^2(\pi\omega\tau_2/2)}
+4\tilde{E}_J\tau_2\\
&\quad=\int\limits_{-\infty}^{\infty}
\left[\frac{\ell_s}{2e_c}\sqrt{1+\frac{\omega_\mathrm{p}^2}{\omega^2}}
\coth\frac{L}{2\ell_s}\sqrt{\frac{\omega^2}{\omega^2+\omega_\mathrm{p}^2}}
-\frac{\ell_s}{2e_c}\frac{2\ell_s}L\,\frac{\omega^2_\mathrm{p}}{\omega^2}
-\frac{\ell_s}{2e_c}\right]
\frac{\pi\,d\omega}{\cosh^2(\pi\omega\tau_2/2)}
+8\tilde{E}_J\tau_2\\
&\quad=g\int\limits_0^\infty\left[
\sqrt{\frac{4\ell_s^2}{L^2}+\frac{1}{u^2}}\coth{u}
-\frac{1}{u^2}-\frac{2\ell_s}{L}\right]du
-g\int\limits_0^\infty\left(\sqrt{1+\frac{1}{u^2}}-1\right)
\tanh^2\frac{\pi\omega_\mathrm{p}\tau_2u}2\,du
+8\tilde{E}_J\tau_2.
\end{align*}
\end{widetext}
The first integral evaluates to $\ln(L/\ell_s)+c_2+O(\ell_s/L)$
where the constant $c_2=c_1+\gamma+\ln{2}-2=-1.567514\ldots$ (here $\gamma=0.577\ldots$ is the Euler-Mascheroni constant). Thus, we can write
\begin{subequations}\begin{align}
&\frac{S_\mathrm{cl}}{g}=\frac{8}\pi\sqrt{\frac{\ell_c}{\ell_J}}
+\ln\frac{L}{\ell_s+\sqrt{\ell_J\ell_c}}
-\Upsilon\!\left(\frac{\sqrt{\ell_J\ell_c}}{\ell_s}\right),\\
&\Upsilon(z)\equiv\int_0^\infty\left(\sqrt{1+\frac{1}{u^2}}-1\right)
\tanh^2\frac{\pi{z}u}2\,du\nonumber\\ &{}\quad\qquad-c_2-\ln(1+z).
\end{align}\end{subequations}
Thus defined $\Upsilon(z)$ is a monotonic bounded function:
\[
1.567514\ldots=\Upsilon(0)\leqslant\Upsilon(z)<\Upsilon(\infty)=1.922\ldots.
\]

\subsection{Functional determinant}

As discussed in Refs.~\cite{Coleman1977, Vainshtein1982}, the tunnelling matrix element~$W$ between two neighboring minima can be represented as
\begin{equation}\label{eq:J}
W=\sqrt{\frac{\Lambda_{j=0}^{(0)}}{2\pi \tau_*}
\prod_{j>0}\frac{\Lambda_j^{(0)}}{\Lambda_j}}\,e^{-S_\mathrm{cl}},
\end{equation}
where $S_\mathrm{cl}$ is the action on the classical instanton trajectory $\vartheta_\mathrm{cl}(\tau)$, found in the previous subsection, $\tau_*$ is defined as \begin{equation}
\frac{1}{\tau_*}\equiv\int_{-\infty}^\infty
\left(\frac{d\vartheta_\mathrm{cl}}{d\tau}\right)^2d\tau,
\end{equation}
while $\Lambda_j$ and $\Lambda_j^{(0)}$ are the eigenvalues of the equation
\begin{equation}\label{eq:Hpsi=Lambdapsi}
\tilde{E}_J\psi(\tau) + \int
K(\tau-\tau')\,\psi(\tau')\,d\tau'+V(\tau)\,\psi(\tau)=\Lambda\psi(\tau),
\end{equation}
for $V(\tau)=-\tilde{E}_J[1+\cos\vartheta_\mathrm{cl}(\tau)]$ and $V(\tau)=0$, respectively. The infinite product in Eq.~(\ref{eq:J}) is over all eigenvalues except the lowest ones, $\Lambda_0=0$ and $\Lambda_0^{(0)}=\tilde{E}_J$. We impose the periodic boundary conditions, $\psi(-\beta/2)=\psi(\beta/2)$, where $\beta\to\infty$ can be viewed as the inverse temperature.

Again, we start with the limiting case $\ell_J\gg\ell_c$, where the classical solution~(\ref{eq:arctan}) yields
\begin{equation}
V\left(\tau\right)=-2\tilde{E}_J\,\frac{\tau_{1}^{2}}{\tau^{2}+\tau_{1}^{2}},\quad
\frac{1}{\tau_*}=\frac{2\pi}{\tau_1}.
\end{equation}
It is convenient to pass to the Fourier space, which is discrete, $\omega_m=2\pi{m}/\beta$, $m=\ldots,-1,0,1,\ldots$, because of the boundary conditions $\psi(\beta/2)=\psi(-\beta/2)$. Thus, we decompose $\psi\left(\tau\right)=\sum_{m}\psi_{m}e^{-i\omega_{m}\tau}$. For $\Lambda\sim\tilde{E}_J\ll\omega_K$ we can use the low-frequency expression~(\ref{eq:Klowfreq}) for $K(\omega)$, then the eigenvalue equation~(\ref{eq:Hpsi=Lambdapsi}) becomes
\begin{equation}\label{eq:mpsim}
|m|\psi_m-\sum_{m'}e^{-(2\pi\tau_1/\beta)|m-m'|}\psi_{m'}
=\frac\beta{2\pi\tau_1}\,\frac{\Lambda-\tilde{E}_J}{\tilde{E}_J}\,\psi_m.
\end{equation}
Let us define a function
\begin{equation}
\chi_m=-\frac{\delta_{m,0}}{1-e^{-2\kappa}}+\theta(m+1/2)\,e^{-\kappa{m}},\quad
\kappa\equiv\frac{2\pi\tau_1}\beta,
\end{equation}
where $\theta(x)$~is the Heaviside step function,
then all eigenvectors and eigenvalues of the problem~(\ref{eq:mpsim}) can be written down explicitly:
\begin{align}
&\psi_m=e^{-\kappa|m|},\quad
\Lambda/\tilde{E}_J=1-\kappa\coth\kappa,\nonumber\\
&\psi_m=\chi_m,\quad\Lambda/\tilde{E}_J=1,\nonumber\\
&\psi_m=\chi_{-m},\quad\Lambda/\tilde{E}_J=1,\nonumber\\
&\psi_m=\chi_{m-1},\quad\Lambda/\tilde{E}_J=1+\kappa,\nonumber\\
&\psi_m=\chi_{1-m},\quad\Lambda/\tilde{E}_J=1+\kappa,\nonumber\\
&\psi_m=\chi_{m-2},\quad\Lambda/\tilde{E}_J=1+2\kappa,\nonumber\\
&\psi_m=\chi_{2-m},\quad\Lambda/\tilde{E}_J=1+2\kappa,\nonumber\\
&\ldots.\label{eqs:Lambda}
\end{align}
At the same time, the eigenvalues of~(\ref{eq:mpsim}) with $V(\tau)=0$ are
\begin{equation}
\Lambda^{(0)}/\tilde{E}_J=1,\,1+\kappa,\,1+\kappa,\,1+2\kappa,\,1+2\kappa,\ldots,
\end{equation}
that is, $\Lambda_{j>1}=\Lambda^{(0)}_{j-2}$. However, the total number of eigenvalues must be unchanged by the potential, that is we must recover $\Lambda_j=\Lambda^{(0)}_j$ for very large~$j$, otherwise the infinite product in Eq.~(\ref{eq:J}) will diverge. Thus, we are obliged to consider high frequencies, where the low-frequency expression~(\ref{eq:Klowfreq}) is no longer valid.

At frequencies $\omega\gg{1}/\tau_1$, the potential $V(\tau)$ is a smooth function of~$\tau$, so we can use the WKB approximation (note that the domains of validity of the asymptotic expression~(\ref{eq:Klowfreq}), $\omega\ll\omega_K$, and of the WKB approximation, $\omega\gg{1}/\tau_1$, overlap). In the WKB approximation we write eigenfunctions as $\psi(\tau)=\exp\left[\pm{i}\int_0^\tau\omega(\tau')\,d\tau'\right]$, then $\omega(\tau)$ should be found from the equation
\begin{equation}\label{eq:WKB}
\tilde{E}_J+K(\omega(\tau))+V(\tau)=\Lambda\;\;\;\Rightarrow\;\;\;
\omega(\tau)\approx\omega_\Lambda-\frac{V(\tau)}{K'(\omega_\Lambda)},
\end{equation}
where $\omega_\Lambda$ is the positive solution of the same equation for $V(\tau)=0$, and $K'(\omega)=dK(\omega)/d\omega$. In the presence of~$V(\tau)$, the quantization condition involves the scattering phase,
\begin{equation}
\int_{-\beta/2}^{\beta/2}\omega(\tau)\,d\tau=
\beta\omega_\Lambda+\frac{2\pi\tilde{E}_J\tau_1}{K'(\omega_\Lambda)}=2\pi{m}.
\end{equation}
This gives $\Lambda=\tilde{E}_J+K(\omega_m)-\kappa\tilde{E}_J$, where $m$ must run over all integers, positive and negative, except $m=0$, in order to match Eq.~(\ref{eqs:Lambda}). Then we can calculate 
\begin{align}
&\prod_{j>0}\frac{\Lambda_j^{(0)}}{\Lambda_j}=\prod_{m\neq 0}\frac {\tilde{E}_J+K(\omega_m)}{\tilde{E}_J+K(\omega_m)-\kappa\tilde{E}_J}\nonumber\\
&\quad\qquad{}\mathop{=}\limits_{\beta\to\infty}
\exp\left[\int_{-\infty}^\infty \frac{\tilde{E}_J\tau_1\,d\omega}{\tilde{E}_J+K(\omega)}\right].
\end{align}
The integral can be calculated by choosing some value $\bar{u}$ such that $\ell_s/\ell_J\ll\bar{u}\ll{1}$ and writing
\begin{align*}
&\int\limits_{-\infty}^\infty\frac{du}%
{2\ell_s/\ell_J+(2e_c/\ell_s\tilde{E}_c)u^2+u^2\sqrt{1+1/u^2}}\\
&{}\approx\int\limits_0^{\bar{u}}\frac{2\,du}{2\ell_s/\ell_J+u}
+\int\limits_{\bar{u}}^\infty
\frac{2\,du/u^2}{2e_c/\ell_s\tilde{E}_c+\sqrt{1+1/u^2}}\\
&=2\ln\frac{\bar{u}\ell_J}{2\ell_s}+\int\limits_0^{1/\bar{u}}
\frac{2\,dy}{2e_c/\ell_s\tilde{E}_c+\sqrt{1+y^2}}\\
&\approx 2\ln\frac{\ell_J}{\ell_s}
-\int\limits_{-\infty}^\infty\frac{(2e_c/\ell_s\tilde{E}_c)\,du}{2e_c/\ell_s\tilde{E}_c+\cosh{u}}\\
&=2\ln\frac{\ell_J}{\ell_s}-\frac\zeta{2}\ln\frac{\zeta+1}{\zeta-1},\quad
\zeta\equiv\frac{e_c}{\sqrt{e_c^2-\ell_s^2\tilde{E}_c^2/4}}.
\end{align*}
Collecting all factors, we obtain
\begin{equation}
W=\sqrt{\frac{2\pi}g}\,\frac{e_l}{\ell_s}
\left(\frac{\zeta-1}{\zeta+1}\right)^{\zeta/4}
\left(e^{-c_1}\,\frac{e_l}{\tilde{E}_JL}\right)^g.
\end{equation}

In the opposite limiting case, $\ell_J\ll\ell_c$, the classical solution~(\ref{eq:arctansinh}) yields:
\begin{equation}
V(\tau)=-\frac{2\tilde{E}_J}{\cosh^2(\tau/\tau_2)},\quad
\frac{1}{\tau_*}=\frac{8}{\tau_2}.
\end{equation}
Now the high-frequency asymptotics~(\ref{eq:Khighfreq}) is sufficient, so the 
eigenvalue equation~(\ref{eq:Hpsi=Lambdapsi}) becomes
\begin{equation}
\left(1-\frac{d^2}{ds^2}-\frac{2}{\cosh^2s}\right)\psi(s)=\lambda\psi(s),
\quad s\equiv\frac\tau{\tau_2},\;\;\; \lambda=\frac\Lambda{\tilde{E}_J}.
\end{equation}
This equation can be solved exactly~\cite{LandafshitzIII}. It has one discrete eigenvalue $\lambda=0$, corresponding to the zero mode, and the continuous spectrum for $\lambda\geqslant{1}$. The reflection coefficient is exactly zero, and the transmission coefficient is a pure phase factor. Namely, the right-traveling solution has the following asymptotics at $s\to\pm\infty$:
\[
\frac{i\sqrt{\lambda-1}+1}{i\sqrt{\lambda-1}-1}\,e^{is\sqrt{\lambda-1}}
\mathop{\leftarrow}\limits_{s\to-\infty}\psi(s)
\mathop{\rightarrow}\limits_{s\to+\infty}e^{is\sqrt{\lambda-1}}.
\]
Together with the periodic boundary condition at $\tau=\pm\beta/2$, it determines the quantization of the eigenvalues:
\begin{equation}
\frac\beta{\tau_2}\,\sqrt{\lambda-1}
+2\arctan\frac{1}{\sqrt{\lambda-1}}=2\pi{m},\quad m=1,2,\ldots.
\end{equation}
The same set of eigenvalues is obtained for left-traveling solutions. Then the determinants' ratio evaluates to
\begin{equation}
\prod_{j>0}\frac{\Lambda_j^{(0)}}{\Lambda_j}=
\exp\left[\frac{2}\pi\int_0^\infty\frac{\ln(1+u^2)}{1+u^2}\,du\right]=4.
\end{equation}
Collecting all factors, we obtain
\begin{equation}
W
=\frac{4\tilde{E}_J}{\sqrt{g}}
\left(\frac{\ell_J}{\ell_c}\right)^{1/4}
\left[\frac{\ell_s+\sqrt{\ell_J\ell_c}}{L}\,
e^{\Upsilon-(8/\pi)\sqrt{\ell_c/\ell_J}}\right]^g.
\end{equation}

\end{document}